\RequirePackage{fix-cm}
\documentclass[natbib,smallextended]{svjour3}       
\smartqed  
\usepackage{graphicx}
\usepackage{xcolor}
\usepackage{amssymb}
\usepackage{lscape}
\usepackage[hidelinks]{hyperref}
 \journalname{ISSI book on TDEs}
\usepackage{ref_macros} 

\newcommand{\gammachap}{Gamma-ray Chapter}
\newcommand{\optchap}{Optical Chapter}
\newcommand{\radiochap}{Radio Chapter}
\newcommand{\echochap}{Echo Chapter}
\newcommand{\wdchap}{White Dwarf Chapter}

\newcommand{\ratechap}{Rates Chapter}
\newcommand{\disrupchap}{Disruption Chapter}
 
\newcommand{\diskchap}{Accretion Disc Chapter}

\newcommand{\emischap}{Emission Mechanisms Chapter}

\newcommand{\chan}{{\em Chandra} }
\newcommand{\xmm}{{XMM-{\em Newton} }}
\newcommand{\xmmns}{{XMM-{\em Newton}}}
\newcommand{\swift}{{\em Swift }}
\newcommand{\swiftns}{{\em Swift}}
\newcommand{\rosat}{{\em ROSAT }}
\newcommand{\mfourteenlong}{{XMMSL2~J144605.0+685735 }}

\newcommand{\mfourteen}{{XMMSL2~J1446+68 }}
\newcommand{\mfourteenns}{{XMMSL2~J1446+68}}
\newcommand{\mseven}{{2MASX~0740-85 }}
\newcommand{\msevenns}{{2MASX~0740-85}}
\newcommand{\msevenlong}{{2MASX~07400785-8539307 }}

\newcommand{\sdsstwlong}{{SDSS~J120136.02+300305.5 }}
\newcommand{\sdsstw}{{SDSS~J1201+30 }}
\newcommand{\swtd}{{SWIFT~J164449.3+573451 }}
\newcommand{\swtdns}{{SWIFT~J164449.3+573451}}

\newcommand{\fluxunits}{{erg s$^{-1}$cm$^{-2}$ }}

\newcommand{\lumunits}{{erg s$^{-1}$ }}

\newcommand{\lumunitsns}{{erg s$^{-1}$}}
\newcommand{\lumUnitsns}{{erg s$^{-1}$}}
\newcommand{\brem}{{Bremsstrahlung }}

\newcommand{\msolar}{$M_{\odot}$ }
\newcommand{\msolarns}{$M_{\odot}$}
\newcommand{\galUnits}{gal$^{-1}$yr$^{-1}$ }
\newcommand{\galUnitsns}{gal$^{-1}$yr$^{-1}$}
\newcommand{\doiurl}[1]{\href{http://doi.org/\detokenize{#1}}{{\detokenize{#1}}}}
\newcommand{\growBH}{\href{http://doi.org/10.1007/11403913_27} {doi:10.1007/11403913\_27}}
\newcommand{\arxiv}[1]{\href{http://arxiv.org/abs/arXiv:\detokenize{#1}}
{arXiv:\detokenize{#1}}}
\begin{document}

\title{X-ray properties of TDEs}

\author{R. Saxton \and
        S. Komossa \and 
        K. Auchettl \and
        P.~G. Jonker
}


\institute{R. Saxton \at
              TPZ-Vega for ESA, XMM-Newton SOC, ESAC,
              Apartado 78, 28691 Villanueva de la Ca\~{n}ada, Madrid, Spain\\
              \email{rsaxton@esa.int}           
           \and
           S. Komossa \at
              Max Planck Institut f\"ur Radioastronomie, Auf dem Huegel 69, 53121 Bonn, Germany\\
              \email{astrokomossa@gmx.de}           
              \and
           K. Auchettl \at
           DARK, Niels Bohr Institute, University of Copenhagen, Lyngbyvej 2, 2100, Copenhagen, Denmark\\
              \email{katie.auchettl@nbi.ku.dk} \\          
            \emph{Present address:} School of Physics, The University of Melbourne, Parkville, VIC 3010, Australia\\  
          \and
          P.~G.~Jonker \at
              SRON, Netherlands Institute for Space Research, Sorbonnelaan 2, 3584~CA, Utrecht, The Netherlands.\\
              Department of Astrophysics/IMAPP, Radboud University, P.O.~Box 9010, 6500 GL, Nijmegen, The Netherlands.\\
              \email{p.jonker@astro.ru.nl}
}

\date{Received: date / Accepted: date}

\maketitle

\begin{abstract}
Observational astronomy of tidal disruption events (TDEs) began with the detection of X-ray flares from quiescent galaxies during the ROSAT all-sky survey of 1990-1991. The flares complied with theoretical expectations, having 
high peak luminosities ($L_{\rm x}$ up to $\ge4\times 10^{44}$ erg/s),
a thermal spectrum with $kT\sim$few$\times10^5$ K, and a decline on timescales of months to years, consistent with a diminishing return of stellar debris to a black hole of mass $10^{6-8}$\msolarns.
These measurements gave solid proof that the nuclei of quiescent galaxies are habitually populated by a super-massive black hole.
Beginning in 2000, \xmmns, Chandra and \swift have discovered further TDEs which have been monitored closely at multiple wavelengths. A general picture has emerged of, initially near-Eddington accretion, powering outflows of highly-ionised material, giving way to a calmer sub-Eddington phase, where the flux decays monotonically, and finally a low accretion rate phase with a harder X-ray spectrum indicative of the formation of a disk corona. There are exceptions to this rule though which at the moment are not
well understood.
A few bright X-ray TDEs have been discovered in optical surveys but in general X-ray TDEs show little excess emission in the optical band, at least at times coincident with the X-ray flare.
X-ray TDEs are powerful new probes of accretion physics down to the last stable orbit, revealing the conditions necessary for launching jets and winds.
Finally we see that evidence is mounting  for nuclear and non-nuclear intermediate mass black holes based on TDE flares which are relatively hot and/or fast.

\end{abstract}

\section{Introduction}
\label{intro}

A key unsolved question in extra-galactic astrophysics in the 70s and 80s was, whether black holes exist at the centers of most or all galaxies. While 10\% of galaxies were Active Galactic Nuclei (AGN) and believed to be powered by accretion onto supermassive black holes (SMBH), the remaining 90\% were quiescent, inactive galaxies with no signs of permanent activity. Did they still harbour black holes at their centers ? If so, how could we actually detect such ``dormant'' black holes ?  
Only in a handful of the closest (quiescent) galaxies, could SMBHs be discerned by their impact on stellar rotation curves \citep{Sargent:1978a}.
In order to tell, whether SMBHs existed not only in AGN, but in the majority of quiescent galaxies,  \citet{Rees:1988a} suggested that, if they disrupted a star and subsequently accreted the matter, then they would reveal their presence by a {\em temporary} X-ray flare, lasting a few months.

The integrated X-ray output of the stellar population of a galaxy is
approximately $10^{39} - 10^{41}$ \lumunits \citep{Fabbiano:1986a} whereas the X-ray flare produced by a TDE emits at a significant fraction of the Eddington luminosity, $10^{42} - 10^{45}$ \lumunitsns \citep[e.g. ][]{Rees:1988a}, and is hence relatively easy to detect
thanks to the high contrast.
While it is straight-forward to distinguish an AGN from a quiescent, in-active galaxy by means of optical spectroscopy (see next section), other X-ray source populations like flare stars, novae, or ULXs embedded in external galaxies are too distant to be resolved from their host galaxies by optical spectroscopy. 
However, they can still be readily distinguished by the X-ray peak luminosity of a TDE, which is
orders of magnitude higher than the luminosity of other X-ray transient phenomena such as Novae, Supernovae, flares stars, accreting Galactic binaries and ULX. 

It was shown that X-ray emission from TDEs should follow the
rate of return of stellar debris to the black hole, which to a
first approximation decays with a 
power-law index of -5/3 
\citep{Rees:1988a,Phinney:1989a,1989ApJ...346L..13E,Lodato:2011a}
such that the luminosity is given by:
\begin{equation}
\label{eq:x:lt53}
L_{X} = A (\frac{t-t_{0}}{1\rm{ yr}})^{-5/3} 
\end{equation}

The first detections of TDE X-ray flares were made with the ROSAT satellite (see Sect.~\ref{sec:rosat}), which found transient sources that displayed the predicted TDE characteristics; a short rise to peak, a steady decline, high peak luminosities, a soft X-ray spectrum, and, importantly, occurred in quiescent, in-active galaxies.
With the advent of Chandra, \xmmns, the Neil Gehrels \swift satellite \citep[hereafter \swiftns;][]{Gehrels:2004a} and multi-wavelength follow-up, differences in the spectra and light curve of each event both compared to AGN and within the class of TDEs detected became more apparent. In particular \swiftns, with its ability to perform high cadence, long-term monitoring of the X-ray and optical/UV light curve and spectrum, has had a major impact in the field of TDEs.

The X-rays generated during a TDE are believed to be produced from the innermost stable orbits of the black hole and experience the strong gravitational field.
This
means they can be used to probe the Kerr and Schwarzschild metrics via precession, quasi-periodic oscillations (QPOs) and reverberation \citep[see the \echochap{} and][]{Rees:1990a}. 

At the time of writing\footnote{In this chapter
we include publications written up until mid-2019.}, the number of X-ray emitting TDEs and X-ray emitting TDE candidates is small enough that each one can be discussed individually, which we do in the following sections. A summary of the properties of each object is given in Table~\ref{tab:tde_summary}.

\section{TDE identification: 
how to distinguish between a quiescent galaxy and an AGN by optical spectroscopy}
\label{sec:x:identify}

Astronomers distinguish between quiescent galaxies and Active Galactic Nuclei (AGN). They are routinely identified and distinguished by means of optical spectroscopy. AGN are galaxies which harbor a permanent accretion disk that emits a strong multi-wavelength continuum including X-rays. This photoionises ambient gas,
and produces strong characteristic, long-lived emission lines. Quiescent galaxies are those which {\em do not harbor a permanent accretion disk} and therefore {\em do not emit any luminous X-rays from their cores}, neither variable nor constant, and
they {\em lack the characteristic optical emission lines }\citep{Osterbrock:1989a}.

The two of them can therefore be distinguished by means of {\em optical spectroscopy}: AGN have characteristic, high-ionization {\em narrow emission lines}, which arise in low-density gas at large nuclear distances (the so-called 'Narrow-line Region'; NLR). The NLR is photoionized by the photons from the {\em permanent} accretion disk, and it 
{\em traces the activity over a long time interval
of 1000s of years}. In contrast, quiescent galaxies {\em lack these characteristic
NLR emission lines}. The majority of quiescent galaxies have no emission lines at all while star-formation galaxies have low-ionization lines, very different
from NLRs \citep{BPT81,Osterbrock:1989a, Kewley:2001a,Kauffmann:2003a}.
Therefore, a crucial and very reliable distinction between an AGN
and a quiescent galaxy is to take an optical spectrum, and determine the presence or absence of the characteristic NLR emission lines. 
Identifying quiescent galaxies by optical spectroscopy has been routinely carried out for decades \citep[see][for a good textbook]{Osterbrock:1989a}. 
Optical spectroscopy to check for a quiescent, inactive host galaxy has therefore been an important step in the identification of any new TDE \citep[e.g.][]{KomossaGreiner99}, following the procedures suggested by theoreticians \citep[e.g.][]{Rees:1988a,Rees:1990a} irrespective of the waveband in which it was discovered \footnote{Note, that broad and narrow optical emission lines can also be {\em temporarily} excited by the TDE itself \citep[e.g.][]{Komossa:2008ar,Wang:2012ar}. However, these lines differ from classical NLRs and can be distinguished if there is more than one post-flare optical spectrum, since they are not permanent but will change and fade away quickly.}.

Finally, we would like to note, that (1) low-mass AGN and (2) AGN in LINERs (low-ionization nuclear emission regions) are more difficult to identify spectroscopically \citep[e.g.][]{Greene:2012a}. 
High S/N optical spectroscopy is required in case (1), while (2) LINER-like 
emission lines, which populate a separate region in diagnostic diagrams, are known to  be powered by several different physical mechanisms; shocks, photoionization by old stellar populations, and/or AGN \citep[e.g.][]{Schulz:1994a}.

\section{ROSAT soft X-ray TDEs}
\label{sec:rosat}

Given the low disruption rate of TDEs (about 1 event every 10$^{4-5}$ years per
galaxy; Sect.~\ref{sec:x:rates}), large-area sky surveys are best-suited to detect these events.
The X-ray mission ROSAT was an ideal experiment to systematically search for, and detect,
TDEs. 
Launched in June 1990, it carried out the first imaging X-ray survey of
the entire sky in the soft X-ray band,  
at energies between (0.1-2.4) keV \citep[e.g.][]{Trumper:1982a}. The ROSAT all-sky survey (RASS) lasted for about a
year and was then followed by an eight-year phase of pointed observations of selected
targets, including deep follow-ups of exceptional sources and transients
discovered during the RASS. ROSAT was equipped with
two prime instruments, a high-resolution imager (HRI) providing 
a spatial resolution
of 5 arcsec, and a positionally-sensitive proportional counter (PSPC) which
allowed broad-band X-ray spectroscopy and achieved a spatial resolution of
25 arcsec on-axis.

Several luminous, high-amplitude X-ray flares from quiescent galaxies, matching remarkably
well the predictions of tidal disruption theory \citep[e.g.][]{Rees:1988a,Rees:1990a}, have first been identified by ROSAT. Four main events were
discovered, from the galaxies NGC5905 \citep{Bade:1996a,KomossaBade:1999a}, RXJ1242-1119 \citep{KomossaGreiner99},
RXJ1624+7554 \citep{Grupe:1999a} and RXJ1420+5334 \citep{Greiner:2000ar}. Among these, NGC5905 and RXJ1242-1119 are the best-covered events
in terms of their multi-wavelength follow-up observations and their long-term
X-ray lightcurves, spanning time intervals of more than a decade, with amplitudes
of decline larger than a factor of 1000 \citep{Komossa:2004ar,Halpern:2004a,Komossa:2005a}. We therefore summarize some main properties of these two events
in the subsections below. 

In addition, the ROSAT archive was used to identify new TDEs after the
end of the mission; either sources which were bright during the ROSAT epoch,
and had faded substantially when re-observed with present X-ray missions \citep{Cappelluti:2009ar,Maksym:2014ar,Khab:2014_rosat},
or events which were found to be bright in later observations, but much fainter or undetected with
ROSAT. The ROSAT database has therefore played an important role in the identification of
most soft X-ray TDEs known today.

\subsection{NGC 5905} 

NGC 5905 was first noticed due to its luminous, soft ($kT = 0.06$ keV) X-ray
outburst with (lower limit on the) peak luminosity in the soft X-ray band of 
$L_{\rm x,peak} = 7 \times 10^{42}$ 
erg/s during the RASS{\footnote{This luminosity is even higher, if a powerlaw model is fit, and if there is absorption from the event's host galaxy.}} \citep{KomossaBade:1999a}. It remained bright
for at least 5 days (the time interval while its position was repeatedly scanned during
the RASS) increasing in luminosity to the observed peak (Fig. 1). Given its
daily coverage for almost a week, it has one of the best-covered early
lightcurves of the non-radio-emitting X-ray TDEs.

\begin{figure}
{\includegraphics[height=8.9cm]{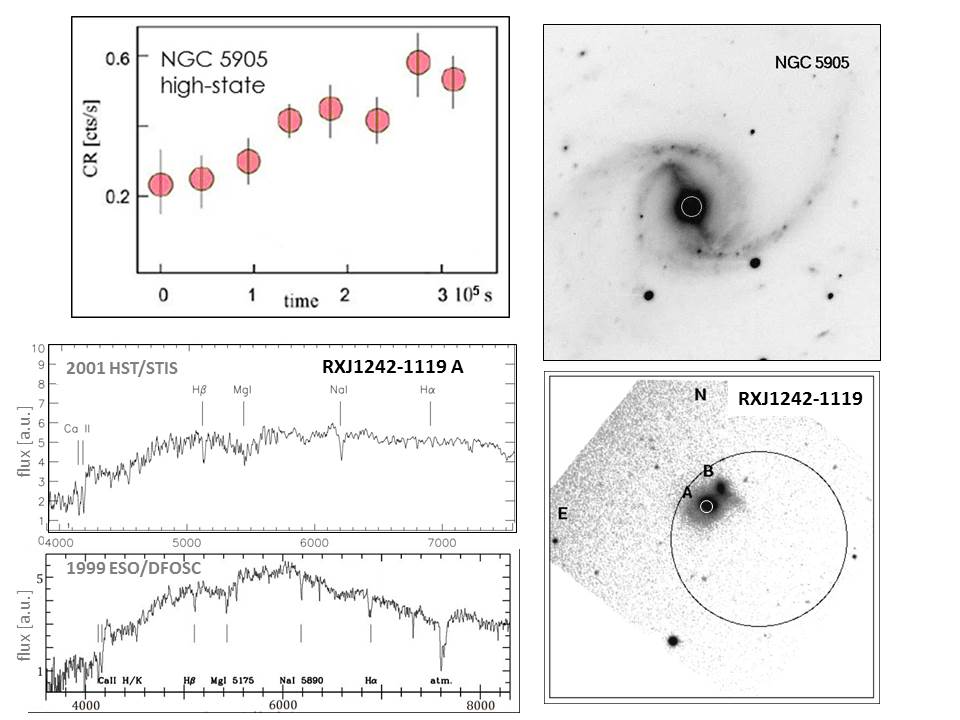}}
\caption{Upper left: Early X-ray lightcurve of NGC 5905 showing its rise to the observed
maximum \citep{KomossaBade:1999a,Bade:1996a}  when  the  galaxy
position was repeatedly covered during the RASS scans.
Upper right: Optical image of NGC 5905 \citep{Komossa:2002a}; a nearby giant barred spiral galaxy. The X-ray flare was observed from the centre of the galaxy. The white circle corresponds to the ROSAT HRI error circle of 5$^{''}$. 
Lower right: Optical image of RXJ1242-1119, adapted from \citet{KomossaGreiner99}. The inital ROSAT X-ray error circle (black) contained two galaxies, both inactive. The X-ray error circle was narrowed down with Chandra (white), which detected fading X-ray emission from the core of RXJ1242-1119A \citep{Komossa:2004ar}. Lower left: Optical spectrum of RXJ1242-1119A [upper panel: HST STIS spectrum from August 2001 \citep{Gezari:2003a}; lower panel: groundbased ESO spectrum from January 1999 \citep{KomossaGreiner99}]. These spectra completely lack emission lines and reveal a quiescent, inactive galaxy.}
\label{fig:x:rosat_results_mosaic}       
\end{figure}

\begin{figure}[t]
 {\includegraphics[height=6.5cm]{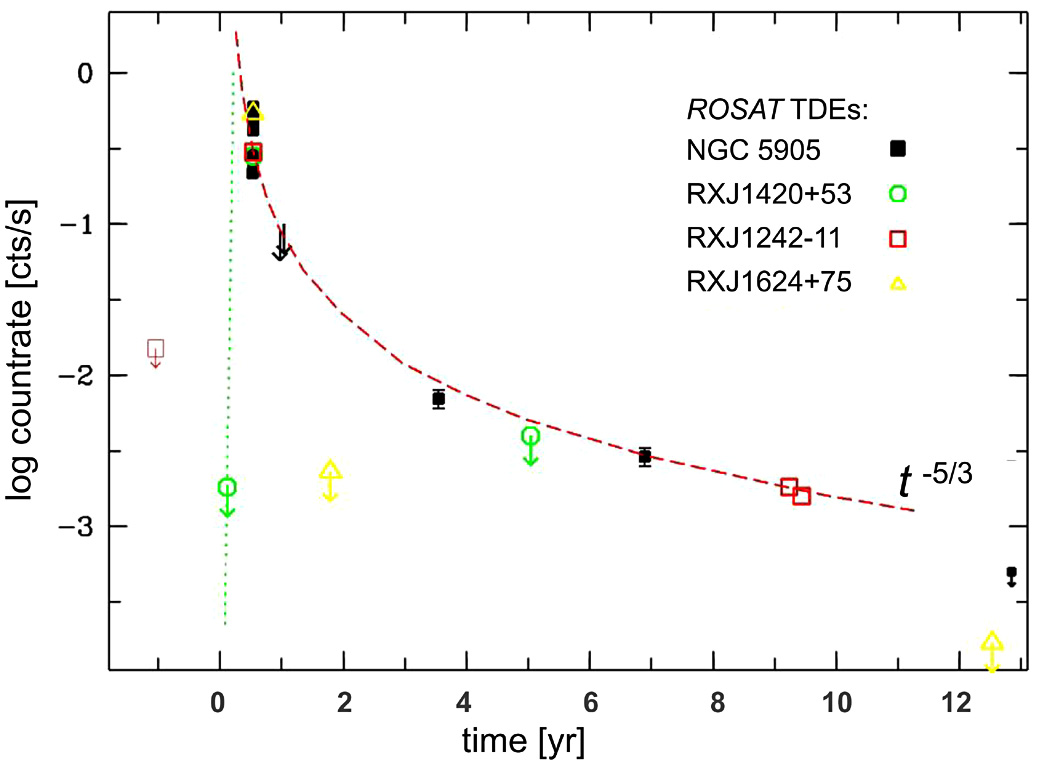}}
\caption{Joint X-ray lightcurve of four TDEs identified with ROSAT, all shifted to the same peak time. The decline is consistent with a $t^{-5/3}$ law (dashed lined). This point was first made based on the ROSAT data of NGC 5905 \citep{KomossaBade:1999a}, and later for the overall luminosity evolution of the sources displayed above \citep[e.g., Fig.1 of][]{Komossa:2004b}. RXJ1242-1119 shows a further drop in X-ray emission at late times (not shown here), deviating from the early phase decline law, implying a total amplitude of decline of a factor $>$ 1000 \citep{Komossa:2005a}. } 
 \label{fig:x:ROSAT_longlight}       
 \end{figure}

X-rays then declined on the timescale of months to years, as observed in multiple re-observations
with ROSAT (Fig. 2). Within the 5 arcsec
positional error (ROSAT
HRI), the X-rays came from the center of this nearby barred spiral galaxy
(z=0.011; Fig.~\ref{fig:x:rosat_results_mosaic}). While the X-ray spectrum was initially very soft, it had hardened
significantly ($\Gamma_{\rm x}=2.4$) 3 years later, when re-observed with ROSAT.
The decline of its X-ray lightcurve is consistent with the predicted $t^{-5/3}$
law (Fig.~\ref{fig:x:ROSAT_longlight}), as first reported based on its ROSAT observations \citep{KomossaBade:1999a} and then confirmed with Chandra \citep{Halpern:2004a}. All observations
of this event are in very good agreement with the predictions \citep{Rees:1990a} from tidal disruption theory
\citep{Bade:1996a,KomossaBade:1999a}.

A very important first step in the identification of every single TDE from a {\em quiescent} galaxy,
is to take an optical spectrum of the host galaxy, and search for the absence of AGN 
activity down to deep limits. 
Ground-based optical spectroscopy before and after the flare has led to a
starburst (HII-type) classification of
the host galaxy of NGC 5905 \citep{KomossaBade:1999a}.
An HST observation after the flare showed
evidence for faint enhanced [OIII] emission from the core of the galaxy
\citep{Gezari:2003a} -- either long-lived or excited by the 
flare itself. 
Using scaling relations
between host galaxy and central SMBH mass for elliptical galaxies \citep{Ferrarese:2000a}, an upper limit on the SMBH mass of NGC\,5905 of a few times 10$^8$\msolar
was estimated; near the upper limit for the tidal disruption of a solar-type
star and as such could imply a spinning SMBH. However, the host of NGC 5905 is a spiral galaxy 
and these tend to lie below the scaling relations
of ellipticals. Using instead the relation of \citet{Salucci:2000a} for spirals gives
a BH mass estimate of $\sim 10^{7}$ \msolar \citep{Komossa:2002a}.

Integrating over its (observed) lightcurve (Eq.~\ref{eq:x:lt53}), only a small fraction of a solar mass needed
to be accreted to power this flare \citep{Komossa:2002a,Li:2002ar}. 
We look at possible explanations for this in Sect.~\ref{sec:x:lowmass}. Note that all such estimates provide a lower limit on the accreted stellar mass, since the events may not have been caught exactly at peak luminosity, and since a significant fraction of the luminosity may be emitted outside the soft X-ray band. 

The morphology of the host galaxy of NGC 5905 is that of a barred spiral
galaxy of type SB; one of the largest spirals known and particularly well resolved in optical
imaging since nearby (Fig.~\ref{fig:x:rosat_results_mosaic}). Its multiple triaxial structures with
a secondary bar might aid the efficient re-filling of its stellar loss-cone orbits,
thereby boosting the stellar tidal disruption rate \citep[see Sect. 3.3.3 of][]{Komossa:2002a}.

\subsection{RXJ1242--1119}

A very luminous X-ray outburst was discovered from the inactive galaxy RX J1242-1119 (at
$z$ = 0.05) with ROSAT during a 'serendipitous' pointed observation. Its X-ray
spectrum was extremely soft, one of the steepest ever identified among
galaxies, and among the steepest in the ROSAT data base, with a photon
index of $\Gamma_{\rm x} = 5.1\pm{0.9}$ \citep{KomossaGreiner99}.
At a soft X-ray luminosity of 4 $\times 10^{44}$ erg/s (a conservative lower limit,
assuming no absorption in the host galaxy itself, and without  applying a
bolometric correction), this event was exceptionally luminous.

Optical spectroscopy of the host galaxy, both ground-based and with the
HST, reveals a quiescent, in-active galaxy, with no emission lines detected
at all down to deep limits \citep{KomossaGreiner99,Gezari:2003a}.

During the RASS itself, RXJ1242-1119 was not detected, implying an initial
amplitude of variability of a factor of $> 20$, and a rise-time of $< 1.5$
years. Deep Chandra follow-ups (cf next section) then revealed the fading
of the X-ray emission from the TDE by a factor of up to $>$1000\citep{Komossa:2004ar,Halpern:2004a,Komossa:2005a} more than a decade after
its high-state. Given its extreme properties and deep follow-ups, this event
continues to be one of the best cases of a TDE identified so far \citep{Komossa:2004ar}.
 
Integrating only over the {\em observed} (0.1--2.4) X-ray lightcurve then gave a strict lower limit on the total emitted energy of $1.6 \times 10^{51}$ erg/s (Eq. 1) and on the accreted mass of 
$ M > 0.01 \eta_{\rm 0.1}^{-1}M_\odot $
for the TDE in RXJ1242-1119 \citep[Sect. 3 of][]{Komossa:2004ar}. 

Based on the host galaxy blue magnitude measured with the optical monitor OM onboard XMM-Newton, $m_{\rm b} = 17.56 \pm 0.05$, the mass of the black hole at the center of the galaxy was estimated, based on the correlation between the absolute blue magnitude of the bulge
of an elliptical galaxy and its SMBH mass
\citep{Ferrarese:2000a}. This yielded an SMBH mass on the order of $M_{\rm BH} \approx 10^8$ M$_\odot$ \citep{Komossa:2004ar}.

\subsection{More recently identified TDEs in the ROSAT archive}
\label{sec:x:ros_archive}

During the ROSAT mission, an X-ray catalogue of $>$100,000 X-ray sources
was created \citep{Voges:1999a}. This data base can still be used, in combination
with the X-ray data bases more recently created with the missions
XMM-Newton and Chandra, to search
for new TDEs bright during the ROSAT observation(s). \citet{Cappelluti:2009ar} identified the X-ray
outburst TDXF1347-3254 ($z = 0.037$) that way, while \citet{Maksym:2014ar} and \citet{Khab:2014_rosat} reported the detection of an
outburst from RBS~1032 ($z = 0.026$).

The event TDXF1347-3254 was the first one identified in a galaxy cluster,
in Abell 3571 \citep{Cappelluti:2009ar}. X-ray emission from one of the cluster galaxies, LEDA 095953, was bright in
ROSAT but decayed by a factor 650 between 1992 and 2004. The black
hole mass, $M_{\rm BH}$, was estimated to be $10^7$ M$_{\odot}$ and the integrated luminosity
was $> 2 \times 10^{50}$ erg/s, implying an accreted mass between 0.001 and 0.03 solar masses. 
The ROSAT PSPC spectrum had an effective $kT=120$ eV.
Multi-band optical/NIR photometry, taken at the time, indicated an inactive host galaxy \citep{Cappelluti:2009ar} which has subsequently been 
confirmed by optical spectroscopy \citep{Wevers:2019b}.

RBS~1032 was a supersoft ($\Gamma_{\rm x} \sim 5$) and luminous ($\sim 10^{43}$ erg/s)
ROSAT PSPC source, which later had faded by a factor $\sim 100 - 300$ when
re-observed with XMM-Newton. It is hosted by an inactive dwarf galaxy
\citep{ghosh:2006a,Maksym:2014ar}. From the shape of the light curve, \citet{Khab:2014_rosat} deduced that the event had been first seen well after peak and that the peak bolometric luminosity was more likely to have been a few $\times10^{44}$ \lumunitsns. This in turn implies that $M_{\rm BH}>10^{6}$ \msolarns. The event remained quite soft ($\Gamma_{\rm x}=3.4\pm{0.3}$) 20 years after discovery.

\subsection{Summary of the properties of the ROSAT TDEs} 
In summary, the ROSAT TDEs are characterized by:
 \begin{itemize}
     \item Soft X-ray (0.1-2.4 keV) peak luminosities up to several 10$^{44}$ \lumunits 
    \item Very soft X-ray spectra near peak, with black-body temperatures in the range $kT_{\rm bb}$ = 0.04--0.12 keV (or, alternatively, with powerlaw indices in the range $\Gamma_{\rm x}=4-5$), followed by a spectral hardening within years.
   \item A decline law of NGC 5905 and RXJ1242-1119 consistent with $t^{-5/3}$, and a drop below that decline law after $t>10$ years. 
   \item Total amplitudes of decline up to factors $>$1000-6000, measured in deep dedicated Chandra follow-up observations (Sect. 4).
   \item Host galaxies, which are optically inactive, radio-inactive, and also X-ray inactive at low-state.
   \item The host galaxy morphology of the nearest event, in NGC 5905, is a giant barred spiral galaxy. 
   \item All of them occurred in the nearby universe ($z=0.011-0.147$). 
   \item Black hole mass estimates are in the range 10$^{6-8}$ M$_\odot$. 
 \end{itemize}

Given their extreme properties, particularly the absence of an optical AGN down to deep limits, huge peak luminosities and total amplitudes of variability (the largest seen in galaxies so far), they continue to be among the most reliably identified TDEs todate.

\section{Using Chandra and XMM-Newton to follow up and detect TDEs}
\label{sec:x:xmm_chandra}

At the end of the 20th century the launch of the XMM-Newton \citep{Jansen:2001a} and Chandra \citep{Weisskopf:2000a} satellites provided the first high spatial and spectral resolution X-ray observation that covered the soft and medium energy bands (0.2-12 keV). \xmm is flying three CCD cameras, with energy
resolution of 50-100 eV at 1 keV and a reflection grating spectrometer (RGS) with a
spectral resolution from 100 to 500 (FWHM) in the energy range 0.33-2.5 keV, as well as an 
Optical Monitor (OM) hosting visible and UV filters. Chandra hosts the ACIS camera, the
High-Resolution camera (HRC) giving sub-arcsecond spatial resolution and low-energy (resolution $>1000$) and high-energy (resolution $=200-800$) transmission gratings.

\begin{figure}[ht]
{\includegraphics[height=5.0cm]{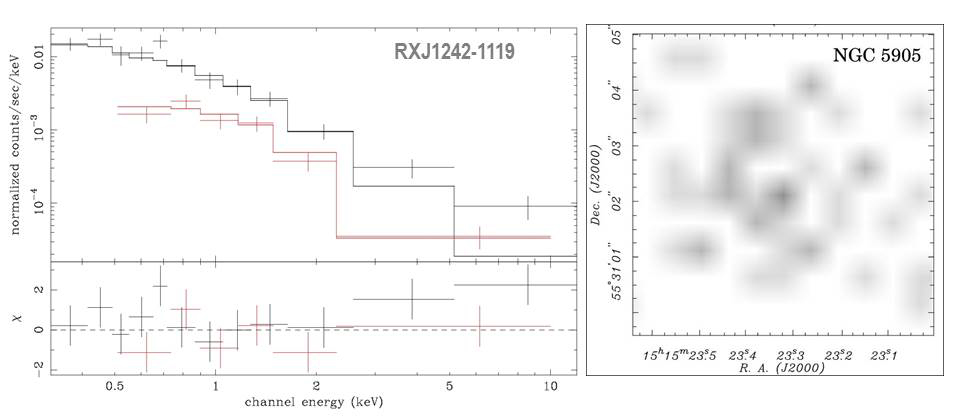}}
 \caption{Left: X-ray spectrum of RXJ1242-1119, the first TDE followed up with XMM-Newton. The observations revealed a strong hardening of the X-ray spectrum compared to the high-state observation \citep{Komossa:2004ar}. 
 Right: Deep Chandra X-ray image of NGC 5905 \citep{Halpern:2004a} 12 years after the flare. Most (or all) of the low-flux state emission is widely extended and from the host galaxy. }
\label{fig:x:ROSAT-followups}       
 \end{figure}

These observatories were used for the first deep follow-up campaigns of the TDEs initially discovered with ROSAT. 
Chandra and XMM-Newton observations of RXJ~1242-1119 detected faint and fading
X-ray emission a decade after the initial TDE high-state consistent with the optical position of the core of the host galaxy within the $\sim$1 arcsecond spatial error of Chandra, supporting the TDE interpretation \citep{Komossa:2004ar}.
With XMM-Newton, a first high-quality post-flare spectrum of the faint late-state X-ray emission from a TDE (RXJ~1242--1119)
was obtained (Fig.~\ref{fig:x:ROSAT-followups}), which showed a significant spectral hardening of the event, from $\Gamma_{\rm x} \sim 5.1\pm{0.9}$ during high-state to $\Gamma_{\rm x} = 2.5\pm{0.2}$ with \xmm \citep{Komossa:2004ar}, possibly related to the formation of an accretion-disk corona, initially absent. We look further at the long-term  evolution of the X-ray spectra of TDEs in Sect.~\ref{sec:x:ltse}. 

\begin{figure}
\label{fig:x:ngc5905_Chandra}    \end{figure}
 
Chandra and \xmm have been used for catalogue searches of TDEs, and also for the quick identification of new TDEs at peak by comparison with the RASS X-ray catalogues. 

XMM-Newton generally makes observations of a few 10s of ks detecting around 50 sources within its quarter square degree field-of-view. These ‘serendipitous’ sources have been collated into a catalogue, at the time of writing in its third incarnation \citep[3XMM;][]{Rosen:2016a} and containing upwards of 500,000 independent sources from 2\% of the sky. Chandra has produced a catalogue \citep[CSC 2.0; ][]{Evans:2010a} with 315,000 sources from 1\% of the sky.

With a mean flux of $10^{-14}$ ($10^{-15}$) \fluxunits for XMM-Newton (Chandra) in the 0.2-2 keV energy band the catalogued sources are generally too faint to be compared with historical observations in a search for variability. However, many fields have been observed on multiple occasions and it is possible to look at the flux history of a significant fraction of the catalogue for up to two decades. These comparisons have yielded a number of candidate TDEs.

\subsection{TDEs discovered using XMM-Newton and Chandra}
\subsubsection{2XMMi~J184725.1-631724}

\begin{figure}
  \includegraphics{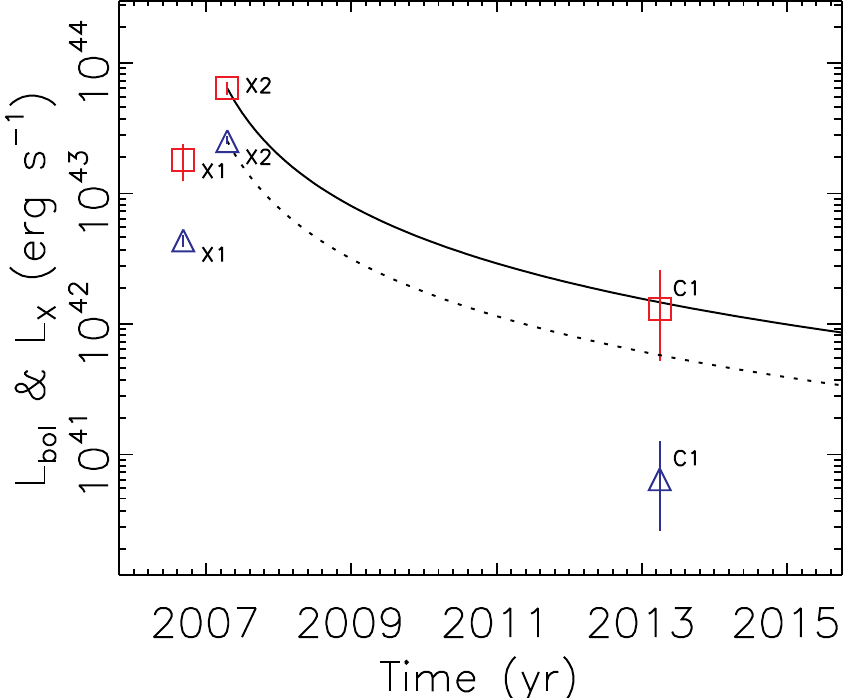}
\caption{The soft X-ray (blue) and bolometric (red) light curve of 2XMMi J184725.1-631724, with X1, X2 and C1 representing the first and second \xmm observations and the first Chandra observation respectively \citep[adapted from][]{Lin_1847_2018}. The solid line plots a
fit to the bolometric luminosity of the function
$(t - t_{0})^{-5/3}$ with $t_{0}$ set at 1 month before the
date of X1, while the dashed line fits the same function to the soft X-ray luminosity.}
\label{fig:x:lc_1847}       
\end{figure}

One such object was 2XMMi~J184725.1-631724 \citep{Lin_1847_2011}, detected in an \xmm pointed observation with a peak observed soft X-ray flux of $2\times10^{-12}$
\fluxunits in 2007 but not detected by a further XMM-Newton observation in 2013 ($F_{X}<2.5\times10^{-14}$\fluxunits), with Chandra confirming a drop in flux by a factor 1000 \citep{Lin_1847_2018} in 2013 (Fig.~\ref{fig:x:lc_1847}). The source whose position is consistent with the nucleus of the optically-inactive galaxy IC~4765-f01-1504 (z=0.0353), was initially flagged due to its very soft spectrum (equivalent black-body temperature of kT$<100$ eV). 

A fainter, but still flux-enhanced, observation of 2XMMi~J184725.1-631724 in 2006 allows the disruption time to be constrained to within about one month enabling the index of the flux decay law to be well measured in this TDE. Surprisingly it turns out to be considerably steeper than -5/3. However, we must bear in mind that this is an in-band flux whereas we should be comparing the bolometric radiation output, under the assumption that this traces the fall back rate of the stellar debris. The peak bolometric luminosity of the TDE was $6\times10^{43}$\lumunits, and the BH mass $M_{\rm BH}=1\times10^{6-7}$\msolarns, based 
on the host galaxy V and K magnitudes, with the peak soft X-ray spectrum dominated by a thermal model of kT$\sim80$ eV. As the spectrum appears to be dominated by black-body radiation, the emission temperature should decrease as $T\propto L^{1/4}$ (see Sect:~\ref{sec:x:spectral_properties}) and hence in the Chandra observation of 2013 will have cooled to $kT\sim35$ eV. At this temperature, much of the flux is shifted out of the X-ray band and the bolometric correction is correspondingly larger than that needed for $kT=80$ eV. Accounting for this effect brings the index of the {\em bolometric} flux decay into good agreement with -5/3 \citep{Lin_1847_2018}.

\subsubsection{3XMM~J152130.7+074916 and 3XMM~J215022.4-55108:  possible TDEs from intermediate mass black holes}
\label{sec:x:imbh}

\begin{figure}
  \includegraphics{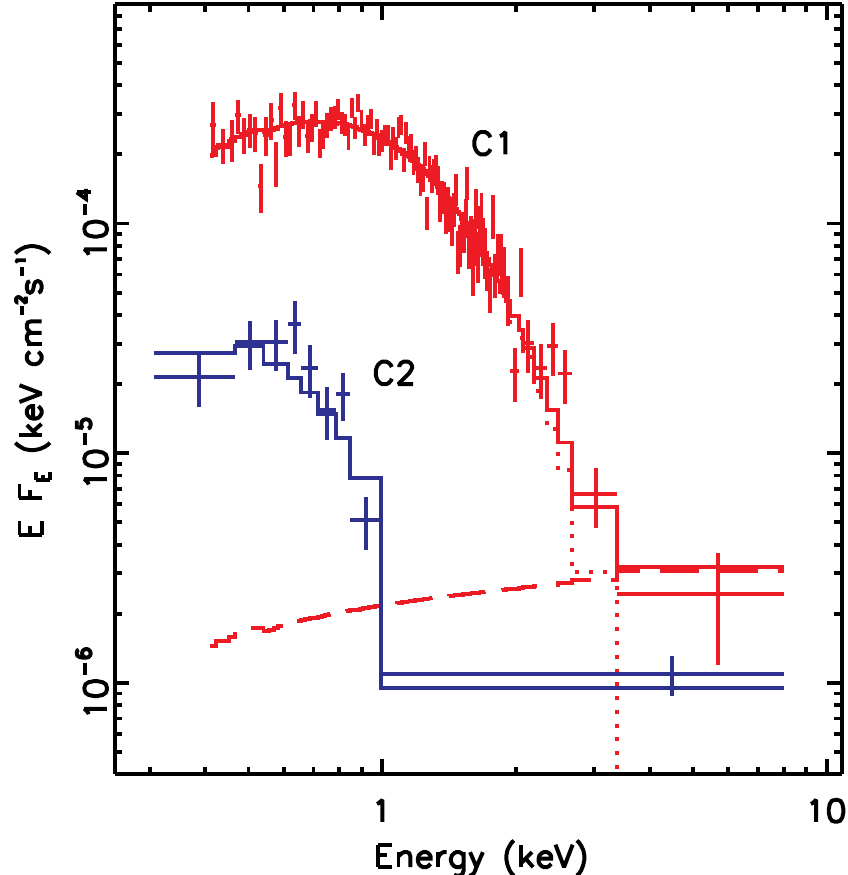}
\caption{Chandra spectra of 3XMM~J215022.4-55108 from 2006-05-05 (red) and 2016-09-14 (blue) fit with a thermal disk model, of $kT=280 (140)$ eV, plus a weak power-law \citep[adapted from][]{Lin_2150}.}
\label{fig:x:2150_spec}       
\end{figure}

The X-ray spectra of TDEs are usually soft in the early phase. If this temperature is related to the black hole mass (see Eq.~\ref{eq:bbody_kt}), then in principle this allows us to identify elusive intermediate mass black holes (IMBH) within the TDE population from their X-ray spectrum, something
that is not possible in AGN surveys where the spectrum is dominated by a power-law
with a mass-independent slope.
The TDE 3XMM~J152130.7+074916 is a good example. Discovered in an \xmm observation in 2000 with a flux 100x higher than that seen in Chandra 4 months earlier, the spectrum could be fit with a disk model with an inner temperature of $kT\sim170$ eV and $M_{\rm BH}$ between 0.19 and 1.4 $\times10^{6}$\msolar depending on BH spin \citep{Lin_1521}. 

An even lower mass was inferred in 3XMM~J215022.4-55108 which lies about 12.5 kpc from the nucleus of a redshift 0.055 galaxy \citep{Lin_2150}. Here the effective temperature peaked at kT=280 eV, reducing to 140 eV as the luminosity decayed from $L_{X}=10^{43}$ to $10^{42}$\lumunits (Fig.~\ref{fig:x:2150_spec}). If these temperatures really come from the inner edge of the accretion disk then the BH in this case has a mass of a few $\times10^{4}$\msolarns. 

This method of estimating black hole mass can only work if the correct spectral model is applied and should be used with caution. For example, a corona (see Sect.~\ref{sec:x:spectral_properties}), if present, would push X-rays to higher energy, artificially increasing the measured temperature and causing the black hole mass to be underestimated. The effect of black hole spin also needs to be taken into account, causing an order of magnitude variation even in simple models \citep[e.g.][]{Lin_1521}.

\subsubsection{TDEs in clusters of galaxies}
\label{sec:clusters}
One 
way of maximising the chances of finding TDEs using Chandra and \xmm pointed observations is to repeatedly observe a rich cluster of galaxies and hence simultaneously monitor many galaxies \citep{WangMerritt:2004a}. \citet{Maksym:2010ar} observed the Abell class 4 cluster, A1689, which at $z\approx0.18$, has a radius of 10 arcminutes nicely fitting into the Chandra field of view. They observed the cluster, which has an estimated 2100 galaxy members, 6 times with Chandra and once with \xmm over an 8 year period. One galaxy, SDSS~J131122.15-012345.6, varied by a factor 30 over the observations, displaying a soft spectrum (kT$\sim100$ eV) and peak $L_{X}>5\times10^{42}$\lumunitsns. From relations with the galaxy magnitude they estimated the black hole mass to be $M_{\rm BH}=1-7\times10^{6}$\msolar.
The light curve could be reasonably well fit with the canonical $t^{-5/3}$ law, from whence the total X-ray luminosity can be found by integrating over Eq.~\ref{eq:x:lt53} 
for the duration of the flare.
They estimate the bolometric correction factor to be 1.4
and found the total luminosity emitted over the event to be $L_{\rm bol}\sim10^{50}$ ergs.
After correcting for the expected emission before the first detection, the equivalent total mass accreted was found from Eq.~\ref{eq:x:mdelta} 
to be $\sim0.01$ solar masses, a surprisingly low value which we address in Sect.~\ref{sec:x:lowmass}.

Another cluster which received a lot of attention from Chandra was A1795, observed a total of 17 times between 1999 and 2010 \citep{Maksym:2013ar}. One dwarf galaxy, WINGS J1348, with a mass of just $3\times10^{8}$\msolar \citep{Maksym:2014br}, showed a bright X-ray source in the first observation which subsequently decayed by a factor $>50$. It appears likely that this flare was first captured by the EUVE satellite, in the 0.016-0.163 keV band, a year before the launch of Chandra, making it the only known TDE detected in the EUV band to date. 
The spectrum in the first Chandra observation was soft ($\Gamma_{\rm x}=4.1$ or $kT=84$ eV) with evidence for further softening over the following year. The sparse light curve again can be reasonably fit to a canonical $t^{-5/3}$ decay index.

\begin{figure}

{\includegraphics[width=12.0cm]{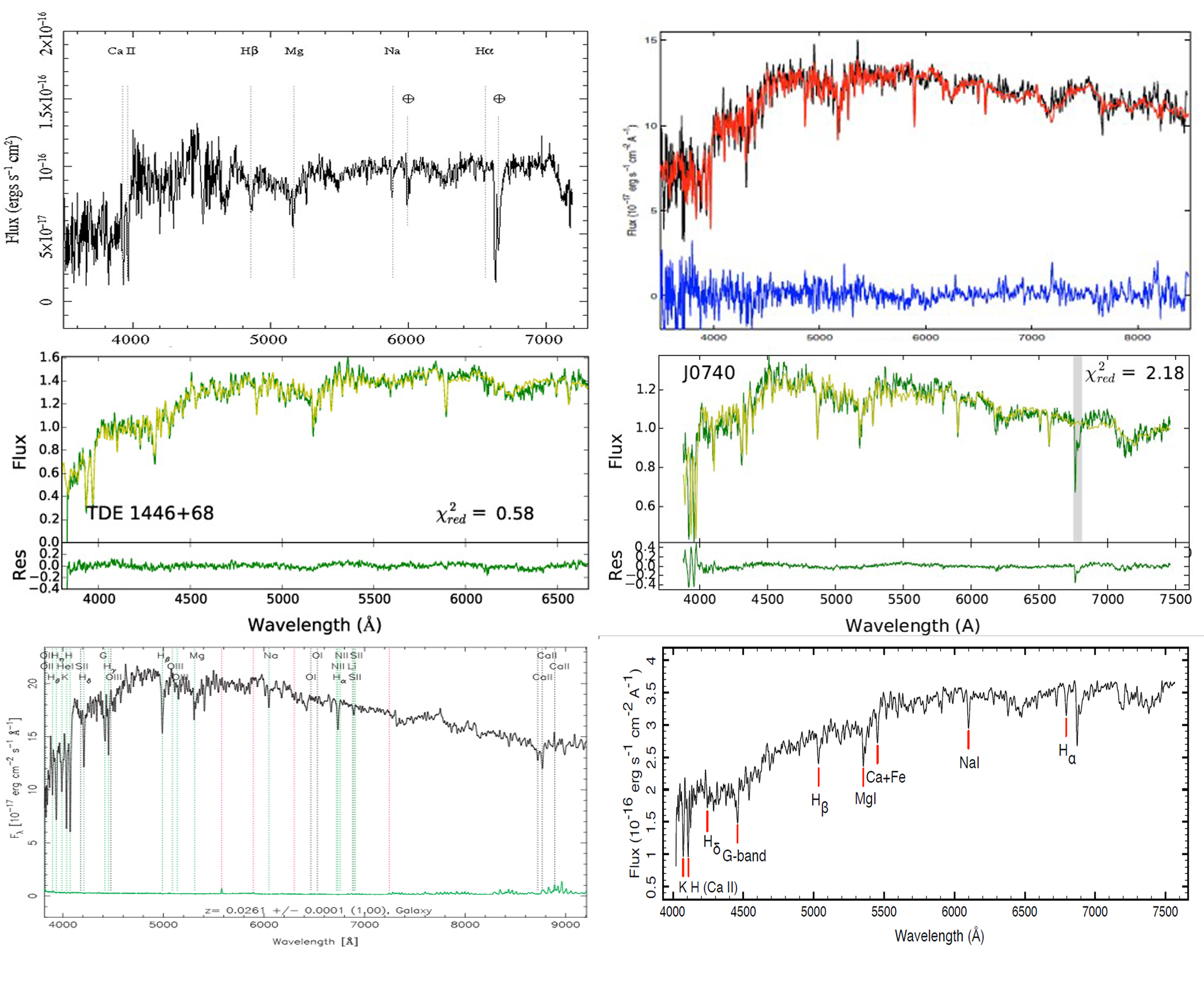}}
\caption{Optical spectra of TDEs detected by \xmmns.
Top left: \sdsstwlong \citep{Saxton:2012a}, top right: SDSS J132341.97+482701.3 \citep{Esquej:2007a}; middle left: \mfourteenlong \citep{Saxton:2019a},
middle right: \msevenlong \citep{Saxton:2017a},
bottom left: RBS~1032 \citep{ghosh:2006a}, bottom right: 2XMMi J184725.1-631724 \citep{Lin_1847_2011}.}
\label{fig:x:optspec}       
\end{figure}

\subsection{TDEs discovered in the XMM-Newton slew survey} 

XMM-Newton slews between targets at a rate of 90 degrees per hour and keeps its most sensitive camera, EPIC-pn \citep{Struder:2001a}, open to record the sky. While a given source passes through the camera in just 15 seconds, the large mirror effective area and short frame time of the detector gives a detection limit of $F_{\rm 0.2-2}=6\times10^{-13}$\fluxunits and a positional accuracy of $\sim8$ arcseconds \citep{Saxton:2008a}. Again at the time of writing, 85\% of the sky has been covered by slews and $>20000$ sources are contained in the XMMSL2 catalogue. The slew sensitivity is well matched with that of the RASS and meaningful comparisons can be made between these surveys to find the bright population of X-ray TDEs.
This was first done by \citet{Esquej:2007a} who found five, previously anonymous, galaxies that showed a factor $>20$ increase in flux between the RASS and slew surveys. Follow-up optical spectra showed that two of these were likely to be due to AGN variability, two (NGC 3599 and SDSS J132341.97+482701.3) were good candidates for TDEs, while XMMSL1~J024916.6-041244 was apparently a persistent Seyfert 1.9 galaxy but showed TDE traits, namely a very soft spectrum and a factor 100 flux decay \citep{Strotjohann:2016a,Auchettl:2017a,Wevers:2019a}.

\subsubsection{\sdsstwlong}
\label{sec:sdss1201}
\begin{figure}
\rotatebox{0}{\includegraphics[height=7.5cm]{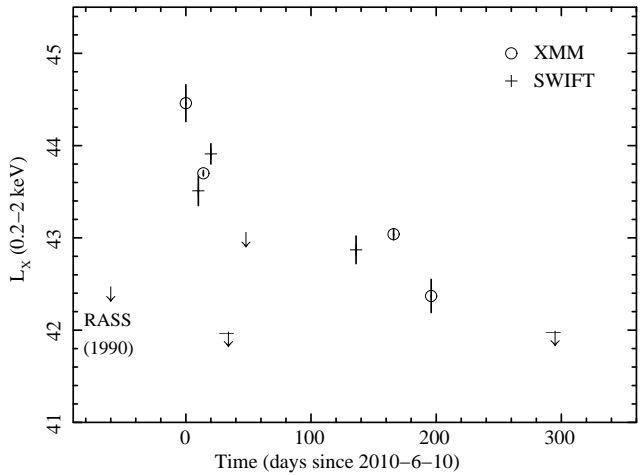}}
\caption{Soft X-ray light curve of SDSS J1201+30. Upper limits are from the
RASS or Swift-XRT \citep[adapted from][]{Saxton:2012a}}
\label{fig:x:lc_1201}       
\end{figure}

The large sky coverage and relatively fast processing of XMM-Newton slews opened up the possibility of investigating events while they were close to the peak of their luminosity. This was first applied to \sdsstwlong (hereafter \sdsstw) spotted in a slew in 2010 \citep{Saxton:2012a,Saxton:2012b}. Fig.~\ref{fig:x:lc_1201} shows the X-ray light curve of \sdsstw as seen in XMM-Newton observations and short exposures taken with \swiftns. Intriguingly the initially high-luminosity emission from this event ($10^{45}$ \lumunitsns) disappeared 3 weeks after discovery, reducing by at least a factor 50 between 2010-06-30 and 2010-07-07. The possibility that this was caused by a temporary local absorption event is unlikely given the long, $>21$ day, duration of the drop in luminosity. 
 Another possibility is that the return of material to the BH was interrupted by the passage of 
 a secondary black hole \citep{Liu:2014a}. This scenario is further explored in Sect.~\ref{sec:x:binarySMBH}.
The emission from this TDE was soft, but wider than a single black-body, being empirically well fit by a \brem model \citep{Saxton:2012a}. During the year of observations, the event got softer showing no evidence for the development of a hard tail.

\subsubsection{\mseven}

\msevenlong (hereafter \msevenns) was detected in a slew in 2014 a factor 20 brighter than an upper limit from ROSAT \citep{Saxton:2017a}. Monitoring by \swift and \xmm over the following 550 days revealed a  drop in X-ray flux by a factor of 70 (Fig.~\ref{fig:x:lc_0740_1446}). 
The host galaxy was shown to be optically inactive (Fig.~\ref{fig:x:optspec}). The first X-ray observations were highly variable, with a flux doubling time of 400 seconds (Fig.~\ref{fig:x:slc_0740}). Using a variability-mass scaling 
relationship \citep{Ponti:2012a} the black hole mass was estimated to be $M_{\rm BH}=3.5^{+6.5}_{-2.4}\times10^{6}$\msolarns. 
Light travel-time arguments then place the size of the emitting region within 73$R_{g}$ of the black hole.

The X-ray spectrum of the event in 2MASS 0740-85 is unusual because it consists of a power-law 
of $\Gamma_{\rm x} \sim 2$ which dominates in the “hard” 2-10 keV X-ray band and contributes 
$\sim25$\% of the total energy. While other TDEs often show a weak hard-tail to the
soft emission, which can be approximated by a (usually) steep power-law \citep[e.g.][]{Lin_1847_2018,2016MNRAS.463.3813H,2018MNRAS.474.3593K} this feature is akin to the dominant power-law
seen almost ubiquitously in AGN \citep{NandraPounds:1994a}. In other sources, \swtdns, Swift~J2058.4+0516 and IGR~J12580+0134 (see the \gammachap{}), the hard X-rays were likely generated in a jet as evidenced by the strong accompanying radio emission. The radio emission from \mseven was rather weak, 1 mJy at 1.53 GHz, and may have been generated from shocks in an outflow rather than a collimated jet \citep{Alexander:2017a}. 

\mseven had a UV flare contemporaneous with the X-ray flare (Fig.~\ref{fig:x:lc_0740_1446}) which decayed with a flatter slope \citep{Saxton:2017a}. 
The relationship between the X-ray and UV emission of X-ray selected TDEs is  
explored further in Sect.~\ref{sec:x:multilambda:uv}.
 
The X-ray and UV emission from \mseven could be reasonably well connected using a multi-coloured disk model (see Sect.~\ref{sec:x:spectral_properties}) suggesting that the wide-band
emission may be coming from a single coherent structure \citep{Saxton:2017a}. 

\subsubsection{\mfourteen}
\begin{figure*}
\begin{center}
\hbox{
\includegraphics[width=0.5\textwidth]{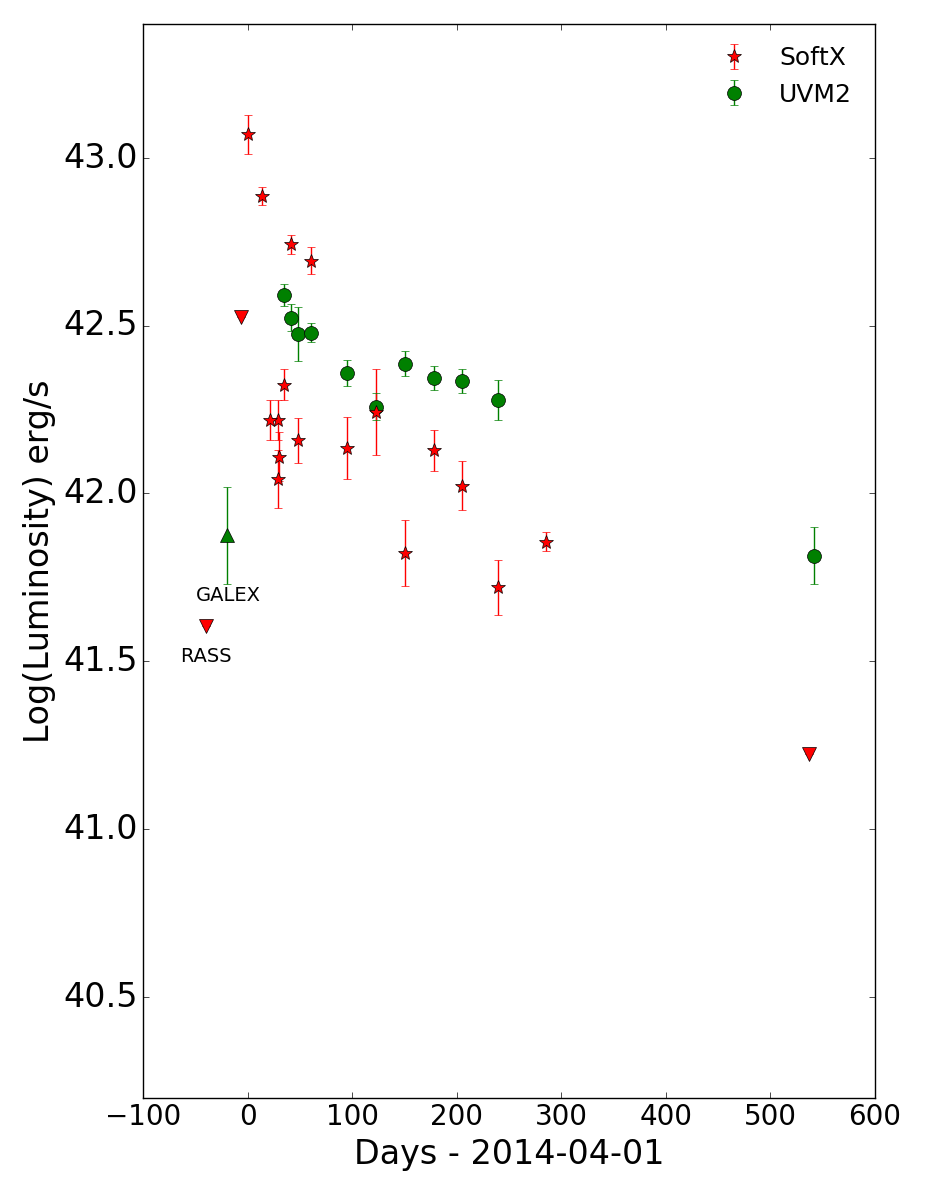}
\includegraphics[width=0.5\textwidth]{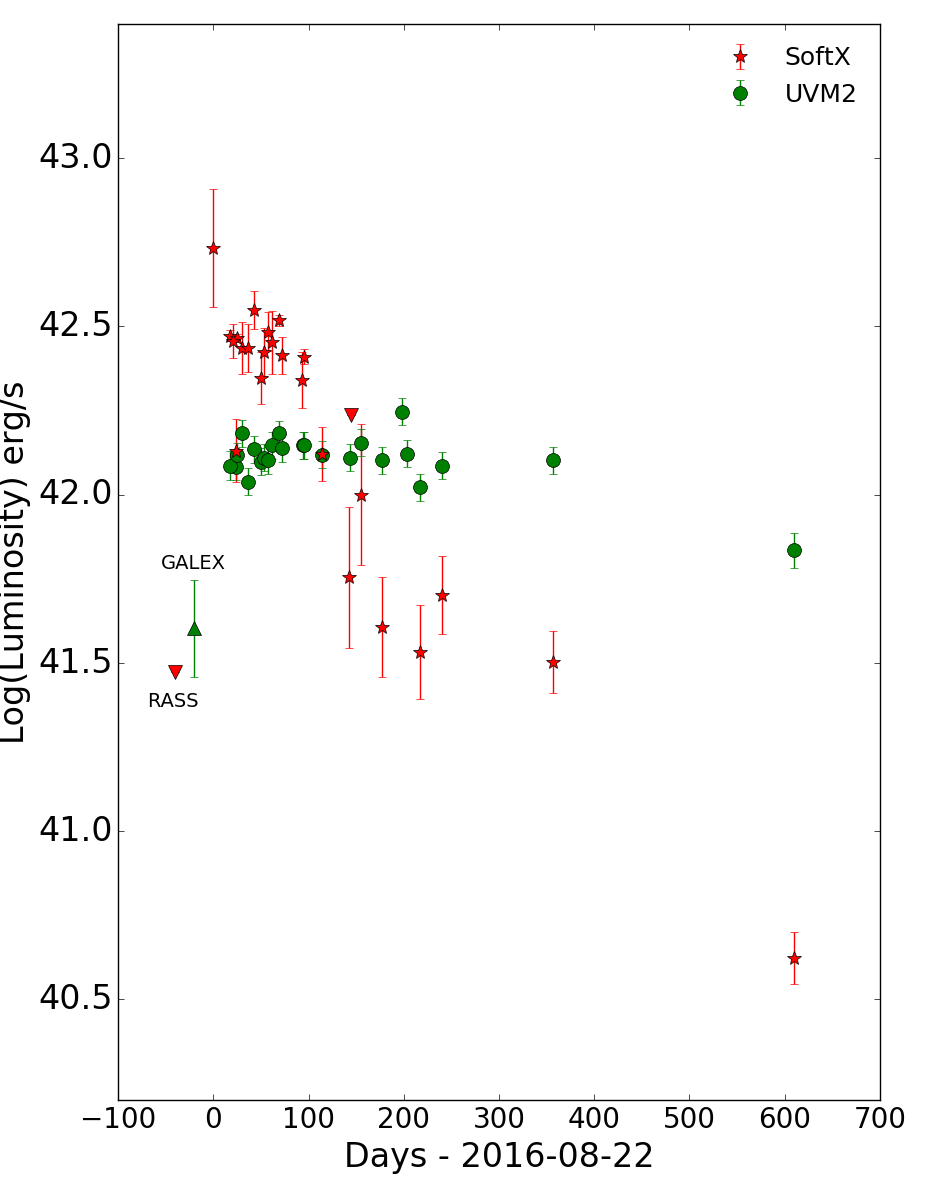}
}
\end{center}
\caption{\footnotesize
The soft (0.2--2 keV) X-ray (red) and UVM2 (2340\AA; green) light curves for
\mseven (left) and \mfourteen (right). Luminosity has been corrected for galactic extinction and includes the contribution from the host galaxy.
GALEX-nuv filter measurements from 2007 have been rescaled to the UVM2 filter in both plots,
which are adapted from \citet{Saxton:2017a} and \citet{Saxton:2019a}.}
\label{fig:x:lc_0740_1446}
\end{figure*}

\begin{figure}[]
    \centering
  \includegraphics[width=0.8\textwidth]{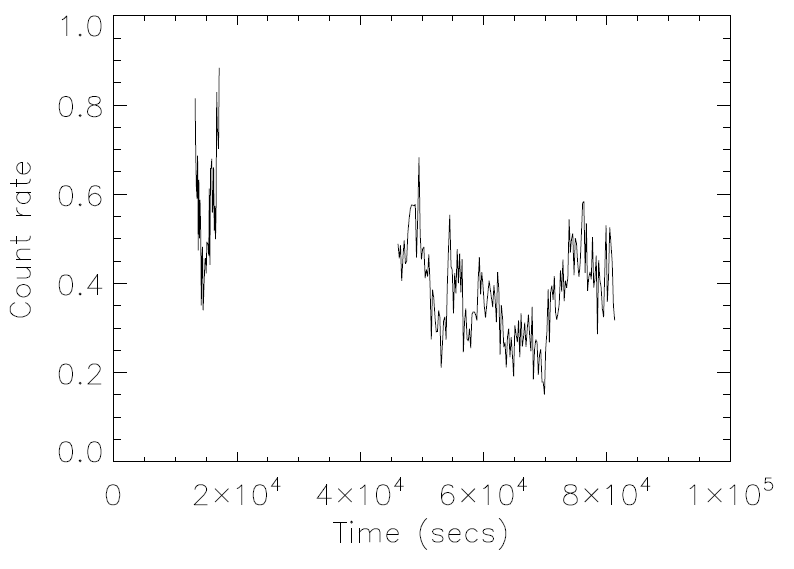}
  \caption{A one-day EPIC-pn light curve of \mseven taken by \xmm soon after discovery,
  showing large short-term variability \citep[adapted from][]{Saxton:2017a}.}
\label{fig:x:slc_0740}
\end{figure}

\mfourteenlong (hereafter \mfourteenns) is another TDE detected in an \xmm slew, this time in 2016. 
In this event the X-ray flux remained stable for the first $\sim100$ days after discovery, before experiencing a drop of a factor 100 over the following 500 days \citep[Fig.~\ref{fig:x:lc_0740_1446};][]{Saxton:2019a}.
The host galaxy is optically inactive (Fig.~\ref{fig:x:optspec}) and with the peak 
bolometric luminosity, $L_{\rm bol}\sim10^{43}$\lumunitsns, interpretations other than a TDE are unlikely \citep{Saxton:2019a}.

It exhibited X-ray spectra consistent with a single power-law, stretching from 0.3--10 keV
with slope, $\Gamma_{\rm x}\sim2.6$. Even in high signal-to-noise spectra there is no significant evidence of a low-energy thermal component in addition to this power-law. The black-hole mass estimated from optical absorption-line widths is quite high, $M_{\rm BH}=7^{+17}_{-5}\times10^{7}M_{\odot}$ \citep{Wevers:2019a}, and thermal emission may be too cool to enter the X-ray band 
(see Sect.~\ref{sec:x:spectral_properties}). The spectral shape remained constant over time in this event, even while the X-ray flux fell by a factor 100.

The event was accompanied by UV emission which remained roughly constant for 400 days before
decaying by a magnitude over the next few hundred days. This disconnect between the UV and X-ray 
flux is unusual and is discussed further in Sect.~\ref{sec:x:multilambda:uv}.

\subsection{X-ray bright events discovered in optical surveys}

For reasons which are currently not well understood, optically and UV-bright TDEs tend to show little or no X-ray emission \citep[][and see the \optchap]{Gezari:2009ar,Arcavi:2014a,Jonker:2019a}.
Nevertheless there are some notable exceptions, two of which are examined below.

\subsubsection{ASASSN-14li}
\label{sec:asassn-14li}

\begin{figure}[]
   \centering
   \includegraphics[width=0.6\textwidth]{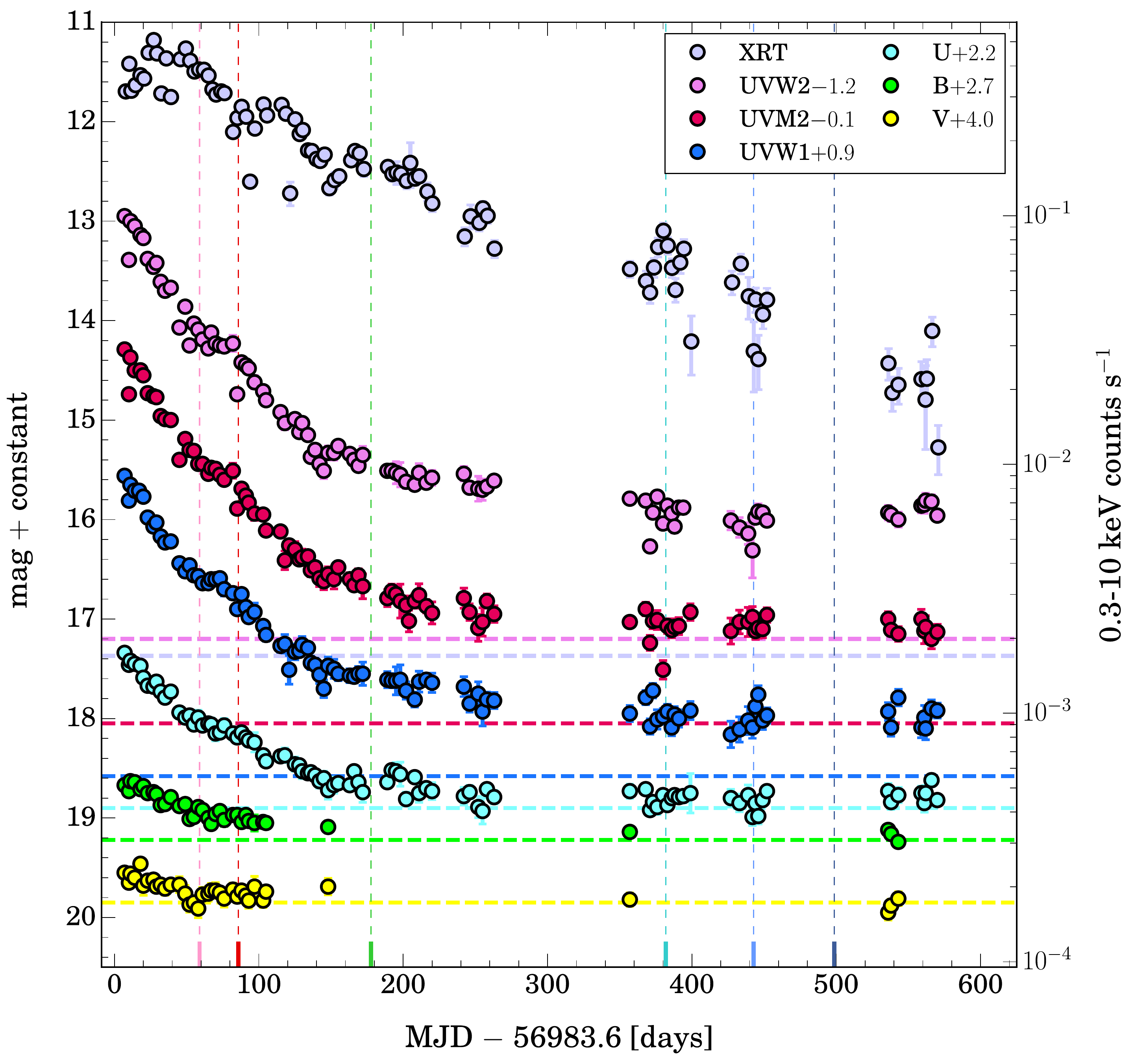}
   \caption{The X-ray, UV and optical light-curves of ASASSN-14li \citep[taken from][]{2017MNRAS.466.4904B}.}
   \label{fig:a14lilc}
\end{figure}

ASASSN-14li was first discovered by the All-Sky Automated Survey for SuperNovae (ASASSN: \citealt{2014ApJ...788...48S}) on 22nd of November 2014. This source was coincident with the center of galaxy, PGC043234 which is located at a distance of $\sim90$ Mpc and its optical spectra exhibit characteristics consistent with that of a TDE (see the \optchap). An extensive multi-wavelength monitoring campaign using \swift revealed in addition to its UV and optical emission, strong X-ray emission arising from this event (see Fig.~\ref{fig:a14lilc}: \citealt{2016MNRAS.455.2918H,2017MNRAS.466.4904B}). \cite{2016MNRAS.455.2918H} found that the X-ray luminosity decayed at a much slower rate than that seen in the optical/UV wavelengths, with X-rays becoming the dominant source of emission approximately 40 days after peak. Using Swift observations spanning 600 days, \citet{2017MNRAS.466.4904B} found that ASASSN-14li remained bright in both UV and X-ray wavelengths even at late times, while the total energy radiated in X-ray and UV/optical was comparable. 

Due to its X-ray brightness, ASASSN-14li was a source of many target of opportunity observations by Chandra and XMM-Newton. Both X-ray grating and CCD spectra showed that the X-ray emission could be described by a simple black-body with a temperature $kT\sim50-60$ eV \citep{2015Natur.526..542M, 2016MNRAS.455.2918H,2017MNRAS.466.4904B,2018MNRAS.475.4011B}, which cooled during the decay \citep{2017MNRAS.466.4904B}. 

Using deep grating spectra taken close to the peak of the flare, \citet{2015Natur.526..542M} found that the black-body emission of ASASSN-14li was modified by absorption from photoionised species of N, O, S, Ar, and Ca. The absorption lines were blue-shifted by
$v_{\rm shift} = -360 \pm 50$ km/s initially, slowing at later times \citep{2015Natur.526..542M}. This led the authors to suggest that the X-ray absorption either arises in strong outflows of highly ionized, low velocity X-ray gas, from a super-Eddington wind or from stellar debris.

More recently, \citet{2018MNRAS.474.3593K} found the early time X-ray spectra exhibit a broad, P-Cygni-like absorption feature around $\sim0.7$ keV, which fades with time. Using photo-ionisation modelling,  they find that this feature is consistent with absorption by OVIII in a very fast (0.2c) and highly ionised outflow. Compared to the low velocity outflow detected by \citet{2015Natur.526..542M}, \citet{2018MNRAS.474.3593K}  suggest that this high velocity component is produced much closer to the black hole. Since the ionisation parameters of these two components are similar, it is possible that the lower-velocity component arises from the fast outflow decelerating as it collides with the debris stream or another medium.

\subsubsection{ASASSN-15oi}
\label{sec:asassn-15oi}

ASASSN-15oi was another X-ray bright TDE discovered by ASASSN on the 14th of August 2015. Located at a distance of 214 Mpc, this source exhibited similar spectral features to other optically detected TDEs (see the \optchap) such as declining strong helium features, black-body emission and a declining light curve \citep{2016MNRAS.463.3813H,2018MNRAS.480.5689H}. However, compared to ASASSN-14li, the X-ray emission from ASASSN-15oi was much weaker. At inital discovery, the detected X-ray emission from the source was lower than an upper limit derived using ROSAT. However, follow-up Swift observations revealed behaviour unseen at the time in any other TDE.\footnote{Similar behaviour has now been detected in AT2018fyk \citep{Wevers:2019b} and AT2019azh \citep{Liu:2019a}.} Typically, the X-ray emission of a TDE decays following a simple powerlaw (see Sect.~\ref{sec:x:lc}),
however, in this case the X-ray emission of ASASSN-15oi brightened by nearly an order of magnitude before fading again \citep{2017ApJ...851L..47G,2018MNRAS.480.5689H}. As the timescale of this increase was approximately one year, \citet{2017ApJ...851L..47G} suggested that the behaviour was the result of delayed accretion due to inefficient circularization of the stellar debris stream, while \citet{2018MNRAS.480.5689H} suggested that this behaviour resulted from material surrounding the accretion disk becoming optically thin to X-ray radiation a few months after discovery. Using XMM-Newton, \citet{2017ApJ...851L..47G} found that the X-ray emission from the source was best described by a black-body plus a weak powerlaw component. While at both early and late times they find the temperature of black-body component does not change significantly ($\sim$45 eV), the powerlaw component may become softer with time ($\Gamma_{\rm initial}\sim2.5$ to $\Gamma_{\rm later} \sim 3.3$). Using the Swift observations, \citet{2018MNRAS.480.5689H} find that the black-body component is an order of magnitude stronger than the powerlaw component, and is responsible for the flux variations.

\subsection{Accreted mass}
\label{sec:x:lowmass}

The mass accreted during an event can be estimated by 
integrating the total emitted X-ray luminosity

\begin{equation}
\label{eq:x:mdelta}
\Delta M = \frac{k_{\rm Bol}}{\eta c^{2}} \int_{t}^{\infty}L_{X}(t) dt
\end{equation}

where $\eta$ is the efficiency of conversion of gravitational 
energy into radiation, generally taken to be $10$\%,
and $k_{\rm Bol}$ is the factor to correct X-ray to bolometric luminosity
\citep[see][for a recent description]{Netzer:2019a}.
From this calculation, the accreted mass often appears to be  $\sim0.01$ solar masses 
\citep[][and references in Sects.~\ref{sec:rosat} and ~\ref{sec:clusters}]{2017MNRAS.466.4904B,Auchettl:2017a,Saxton:2017a};
a low value which potentially leads to a missing energy problem \citep{Piran:2015br}.

We expect from stellar population analysis that the average mass of a disrupted main sequence star will be between 0.1 and 1 solar mass \citep[e.g.][]{2014ApJ...783...23G,vanVelzen:2019b,2018arXiv180108221M}. If the fraction of stellar debris returning to the black hole is $\sim20-50$\%, as dynamical studies predict \citep{Ayal:2000a,Bonnerot:2019a} then, at face value, $\lesssim 10\%$
of the available accretion energy is being converted into radiation in these events. There are several possible explanations for this: 
\begin{itemize}
\item	These flares are actually caused by the stripping of the atmosphere of an evolved star \citep{MacLeod:2012ar} and hence much less material is available for accretion.
\item The conversion of mass to light factor ($\eta$) is around 0.01, i.e. just 10\% of that seen in AGN.
\item We always miss the peak of the emission and underestimate the total luminosity. 
\item The initial super-Eddington accretion produces strong winds which push a large fraction of the material away from the black hole \citep[e.g.][]{Metzger:2016b,2018ApJ...859L..20D}.
\item A large fraction of the radiation is actually emitted in the unobservable EUV and the bolometric correction is seriously underestimated \citep[e.g.][]{Komossa:2004ar}.
\item The returning matter forms an accretion structure which evolves viscously and hence drains into the black hole at a much lower rate than $t^{-5/3}$ \citep{Cannizzo:1990ar,vanvelzen:2019a}
\item We are underestimating the column density of material surrounding the event \citep{Auchettl:2017a}.
\item A combination of the above or a further unknown factor.
\end{itemize}

\subsection{X-ray spectra} 
\label{sec:x:spectral_properties}
The spectral features imprinted by the environment of an AGN are constant for decades, even if the details may vary on timescales as short as minutes. These are well described in e.g. \citet{Netzerbook} and in essence consist of a pseudo-power-law, dominant from 2-100 keV, with a slope of roughly 2 \citep{TurnerPounds:1989a,NandraPounds:1994a}, a soft emission component, which may be due to Compton-upscattering by cool electrons \citep[$kT_{e}\sim0.2$ keV;][]{Done:2012a} of photons generated in the disk and a pure thermal soft X-ray component, usually hidden by the stronger Comptonised component.This emission is modified by reflection off distant neutral material and off the ionised disk and various absorption features with various ionisation states and column densities. In radio-loud objects a jet produces very strong power-law emission which tends to mask the other features. This complex mix requires very high signal-to-noise spectra to begin to deconvolve the individual components.

The prediction is that in the simplified case of a TDE, where we add material at a well-determined date to a SMBH, we ought to be able to determine the temporal onset of each of the components mentioned above and hence achieve a better understanding of persistent accreting SMBH
and the timescales involved in the distribution of matter about the black hole. 

A wide range of empirical models have been used to fit the broad-band \xmm and Chandra spectra of TDEs. A single-temperature black-body, with luminosity given by

\begin{equation}
    L=\sigma AT^{4}
\end{equation}
where $\sigma$ is the Boltzmann constant, $A$ is the surface area of the emitting region and T is the temperature in Kelvin, is a reasonable description in a number of cases (e.g., early emission of ASASSN-14li; Sect~\ref{sec:asassn-14li}).
As the stellar debris returns to the BH it should form a hot, optically-thick structure, with an effective black-body temperature dependent on the BH mass and the mass accretion rate, expressed in terms of the Eddington accretion rate, 
$\dot{M}_{\rm Edd}=L_{\rm Edd}/\eta c^{2}$ \citep{1999ApJ...514..180U}

\begin{equation}
\label{eq:bbody_kt}
kT \sim 40 M_{6}^{-1/4} \dot{M}_{\rm Edd}^{1/4} \qquad  (eV)
\end{equation}

where $M_{6}$ is the mass of the black hole in units of $10^{6}$\msolar
and $\dot{M}_{\rm Edd}$ is the Eddington-limited accretion rate.

\begin{figure}[]
  \centering
  \includegraphics[width=0.6\textwidth]{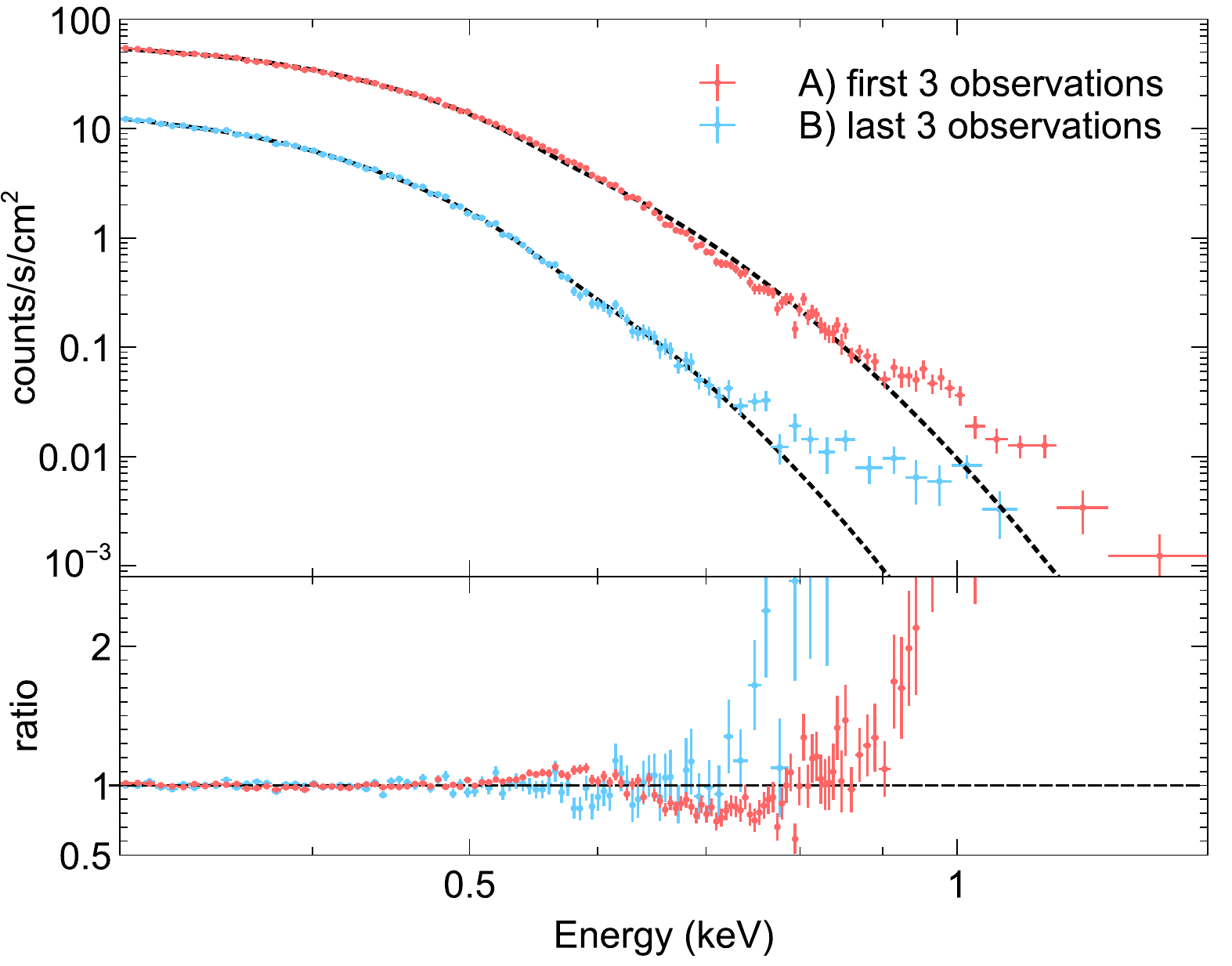}
  \caption{X-ray spectrum of ASSASN-14li from early (red) and late (blue) \xmm observations. Figure adapted from  \citet{2018MNRAS.474.3593K}. The best fit with a single black-body is shown as a dotted black line. A hard excess is evident in both fits.}
  \label{fig:x:spec_assasn14li}
\end{figure}

However, these fits often leave a high-energy tail 
(e.g. 2XMMi J184725.1-631724: \citet{Lin_1847_2011}, SDSS~J1201+30: \citet{Saxton:2012b}, 
ASASSN-14li: \citet{2018MNRAS.474.3593K}, 3XMM~J150052.0+015452: \citet{Lin:2017a}, ASASSN-15oi: \citet{2018MNRAS.480.5689H})
which needs further explanation.
A high-statistic example is shown in figure~\ref{fig:x:spec_assasn14li}. Several explanations have been offered for this in the literature which we run through below.

\textbf{A: A multi-temperature structure:} This assumes that the material is in a thin-disk configuration, where the emissivity index of the material is strongly weighted towards the centre so that most of the emission occurs at the higher temperatures. A common implementation is the {\em diskbb} model \citep{Makishima:1986a}. The range of temperatures produces a broader spectrum than a single-T black-body and can fit the observed spectra in some cases \citep{Lin_1847_2011,Lin_1521, Lin_2150}. This model leads to a high inner disk temperature and, from Eq.~\ref{eq:bbody_kt}, a correspondingly lower  black-hole mass. 

\textbf{B: An empirical power-law:} A single power-law usually proves to be a poor-fit when used to model the whole of a high-statistic TDE spectrum. When it is used, then the slope is usually very steep,
($\Gamma_{\rm x}>4$; see Sect.~\ref{sec:rosat},~\ref{sec:x:xmm_chandra} and~\ref{sec:xreproc})
, in excess of the 
index value of 1.5 to 2.3 \citep{TurnerPounds:1989a,NandraPounds:1994a} ubiquitously found in persistent AGN and believed to be produced by Comptonisation of disk photons by a high temperature ($kT_{e}>100$ keV), optically thin plasma. 
It is more successful in fitting the {\em hard-excess} in good quality spectra of TDEs, but even here the slope tends to be steep (e.g. $\Gamma\sim3.5$ in 2XMM 1847 \citep{Lin_1847_2011} and also in ASASSN-15oi \citealt{2017ApJ...851L..47G,2018MNRAS.480.5689H} or those summarised in \citealt{Auchettl:2017a}). 

\textbf{C: Bremsstrahlung:} This is emission from an optically-thin plasma which produces a wider spectrum than a black-body. It was a good fit to SDSS~J1201+30 with a temperature of 390 eV,
reducing to 280 eV in a later observation \citep{Saxton:2012a}. The main draw-back of this model is that, being optically-thin, at this temperature it should produce substantial narrow line X-ray emission from N, O, Fe \citep[e.g.][]{Mewe:1985a} which was not seen in SDSS~J1201+30 or in other TDEs. 

\textbf{D. Inverse Compton:}
A nearby population of electrons, with significant kinetic energy compared to the photon energies, will give energy to a fraction of the photons in a process known as Compton-upscattering, thus creating a hard tail to the thermal spectrum. This mechanism has been proposed for the soft-X-ray excess regularly seen in the spectrum of AGN
\citep[e.g. ][]{Done:2012a}. 
In the soft X-ray regime, the final energy of an upscattered photon in collision with a thermal electron population of
temperature $kT_{e}$, is related to its initial energy by 
\begin{equation}
E_{f} \sim (1 + \frac{4 kT_{e}}{mc^{2}}) E_{i}
\end{equation}
The hard excess produced by the Inverse Compton effect has a shape which is well approximated by a power-law with slope dependent on the temperature and optical depth of the electrons \citep[see ][]{Nishimura:1986a}. This model provides a more physical fit to SDSS 1201+30 \citep{Saxton:2012b} and has been used to fit the excess in \xmm and Chandra spectra of TDEs such as 3XMM~J150052.0+015452 \citep{Lin:2017a}.
Two popular spectral models are \citep[compbb;][]{Nishimura:1986a} and \citep[comptt;][]{Titarchuk:1994a}.
In these models the typical hard-excess slope of $\Gamma=3-4$ is
provided by a Maxwellian electron population of temperature, $kTe\sim5-10$ keV and optical depth 
$\tau\sim1$.

\subsubsection{Hard power-law emission}

Another emission feature which has been seen is a hard ($\Gamma\sim2$) power-law, equivalent to the dominant emission mechanism in AGN. In some cases this is associated with a relativistic jet (see the \gammachap) accompanied by strong emission at radio wavelengths. In IGR~J12580+0134\footnote {Note that the classification of IGR~J12580+0134 as a TDE has been questioned based on its WISE colours, pre-flare data and hardness ratio evolution \citep[see A.17 of ][]{Auchettl:2017a}} and \mseven the power-law is dominant, even at low energies, but radio emission is modest ($\sim1$ mJy or $L_{R}=10^{37-39}$\lumunitsns). This component seems to be the analogue of the X-ray power-law seen in radio-quiet AGN
\footnote{although \citet{Irwin:2015a} make a case for the 2--10 keV emission coming from the inverse Compton component of the jet in IGR~J12580+0134.},
which is generally explained by Compton-upscattering of disk photons by a population of very high energy electrons (kTe$>100$ keV) located at a few $R_{g}$ from the black hole, a proximity which leads to variations on time scales of minutes to hours. In fact, it is this fast variability, also seen in \mseven (Fig.~\ref{fig:x:slc_0740}), which locates the X-ray emission close to the BH.

\subsubsection{Absorption features}
In the spectra of some events, a good fit can only be obtained by adding one or more {\em ionised} absorption features. This is the case in the CCD spectra of 3XMM J1521 and SDSS~J1201+30 which show evidence for absorption features from apparently outflowing material \citep{Saxton:2012a,Lin_1521}. A more detailed analysis of this  material can be found from the grating instrument on-board \xmmns, the RGS, in the bright TDE ASSASN-14li \citep[see Sect.~\ref{sec:asassn-14li} and ][]{2015Natur.526..542M,2018MNRAS.474.3593K}. Outflows have been regularly found in BHs accreting close to the Eddington limit, in solar mass systems as well as SMBH \citep{Arav:1994a,Crenshaw:1999a,Pounds:2003a}.

Any {\em neutral} absorption ($N_{H}$), in excess of that of our galaxy and the TDE host galaxy, found 
in TDE spectra would be crucial for
constraining the geometry of the debris and accreting material. For example, the reprocessing model
\citep{2018ApJ...859L..20D,Metzger:2016b}  seeks to explain the difference between optical and X-ray TDEs in
terms of a viewing angle (see the \emischap).
The very soft spectra of X-ray selected TDEs are highly sensitive to cold absorption and 
their very presence argues against large columns of neutral gas in the line of sight \citep[e.g.][]{KomossaBade:1999a}. As an example, a TDE at z=0.05, with a Galactic column, $N_{H}=1\times10^{20}$ cm$^{-2}$ and black-body emission of $kT=60$ eV has its 0.2--2 keV flux reduced by a factor 6 by an absorption of $N_{H}=1\times10^{21}$ cm$^{-2}$ intrinsic to the host galaxy of the event.
In many detailed fits, neutral absorption in excess of that of
our own galaxy is not required \citep[e.g. ][]{Saxton:2017a,Saxton:2012a,Lin_1521,Lin_2150}. There are exceptions: \swtd \citep[$N_{H}=2\times10^{22}$cm$^{-2}$;][]{Burrows:2011a}\footnote{note the highly reddened, absorbed host galaxy of this event}; IGR~J12580+0134 
\citep[$N_{H}=7\times10^{22}$cm$^{-2}$;][]{Nikoajuk:2013a}, 3XMM~J150052.0+015452 \citep[$N_{H}=4.2\times10^{21}$cm$^{-2}$;][]{Lin:2017a} and ASSASN-14li
\citep[$N_{H}=1.4\times10^{20}$cm$^{-2}$;][]{2015Natur.526..542M}, but in none of these cases has the excess neutral absorption been seen to change between
observations. This either means that the absorption comes from the host galaxy,
or that the material was produced by the TDE but in a form which did not decrease
in density and was not significantly ionised by the nuclear radiation, during
the event.

The derived absorbing column is of course dependent on the emission model used in the spectral fit. 
\citet{Auchettl:2017a}  
fitted a large sample of TDE spectra
assuming a single absorbed power-law model,
finding that a large fraction of X-ray 
TDEs have column densities ($N_{H}$) that are at least two times greater than the 
Galactic column density measured along the line of sight to these events. This result  holds when a power-law is the correct emission model for the TDE and is useful for investigating variations with time. It does not return the correct absolute value though, for the different spectral models which are commonly seen in TDEs, e.g. the multi-component spectrum of ASASSN-14li \citep[Fig.~\ref{fig:x:spec_assasn14li} and][]{2015Natur.526..542M,2018MNRAS.474.3593K}. 

\subsubsection{Long-term spectral evolution}
\label{sec:x:ltse}

It is interesting to see what happens to the accreting material after many years when the accretion rate has dropped well below the Eddington limit. 
The late-time spectrum of the ROSAT TDE RXJ~1242-1119 was measured with XMM-Newton and implied a strong hardening of the X-ray spectrum 
from $\Gamma_{\rm X}=5.1$ to $\Gamma_{\rm X}=2.5$
\citep{Komossa:2004ar}.
A dedicated study of the late-time spectra of ROSAT TDEs by Chandra, taken $>10$ years after the disruption, showed relatively hard spectra ($\Gamma<2.5$) in all cases  \citep{Halpern:2004a,Vaughan:2004a}. 
The Chandra study is complicated by the very low statistics  ($<25$ photons in each spectrum) and by the low-luminosity of the residual emission ($L_{X}=5\times10^{39}$ to $2\times10^{41}$\lumunits) which is comparable to that seen from the binary population of the galaxy, at least in the case of NGC 5905, where most or all of the low-state emission is extended and clearly not associated with the nucleus \citep{Halpern:2004a}. In NGC 5905 a spectral hardening was already measured with ROSAT itself, 3 years after the maximum \citep[$\Gamma_{\rm X}=4.0$ to $\Gamma_{\rm X}=2.4$;][]{KomossaBade:1999a}.
There is some theoretical expectations that accreting debris will collapse into a thin disk leaving a long-lived low-level emission for many years. Such emission has accretion rates of $\dot{M}\sim10^{-4}$ and an index  $\Gamma_{\rm X}\sim2$ \citep{Cannizzo:1990ar}.
In the case of RXJ1242-1119, the latest deep Chandra follow-ups have shown a deviation from the early-phase decline law, in form of a deep dip below the early-phase decline law, and may indicate a change in accretion mode \citep{Komossa:2005a}, as predicted by e.g. \citet{Rees:1990a}.  

In NGC 3599 the spectrum was still relatively soft 6 years after the peak emission \citep{Saxton:2015a} with an equivalent fit of a power-law of $\Gamma=2.7\pm{0.3}$ plus a kT=50 eV black-body.  In RBS 1032 the spectrum after 20 years had hardened from 
$\Gamma=5$ to $\Gamma=3.4$ \citep{Maksym:2014ar}.
Interestingly, \citet{Jonker:2019a} found late-time X-ray emission in three optically-selected TDEs. The spectra of these could be modeled with power-laws of $\Gamma=2.5-3.9$, compatible
with those of the X-ray selected events. 

The flux of the soft black-body component certainly drops over the years but due to the paucity of high
quality early and late-time spectra, it is not yet clear whether the harder component remains roughly constant or increases over time.
More observations, with higher statistics, are needed to decipher what the long-term spectral hardening means in physical terms.

\section{Light curves}
\label{sec:x:lc}
\begin{figure}
	\begin{center}
		\includegraphics[width=0.55\textwidth]{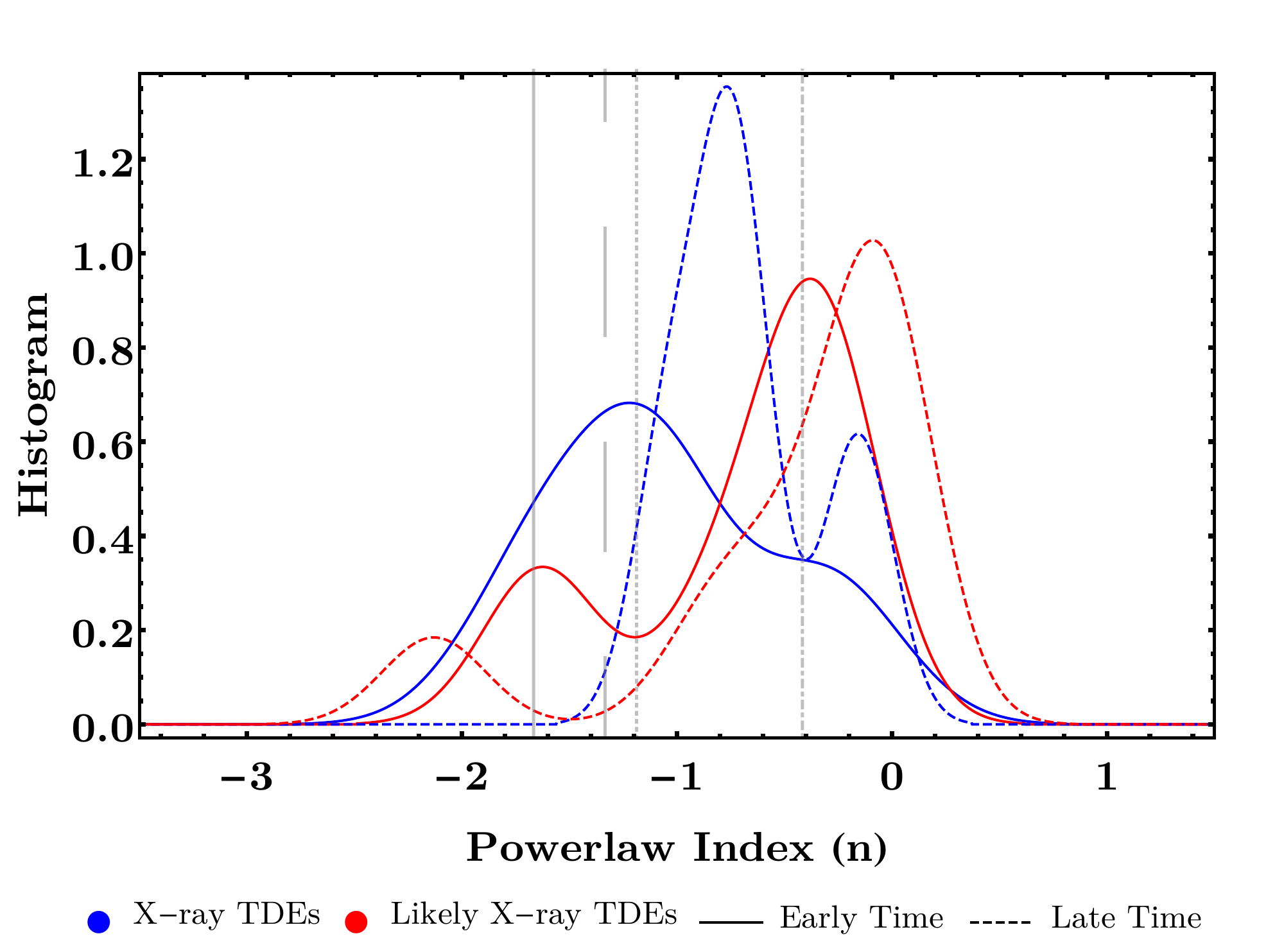}
		\includegraphics[width=0.55\textwidth]{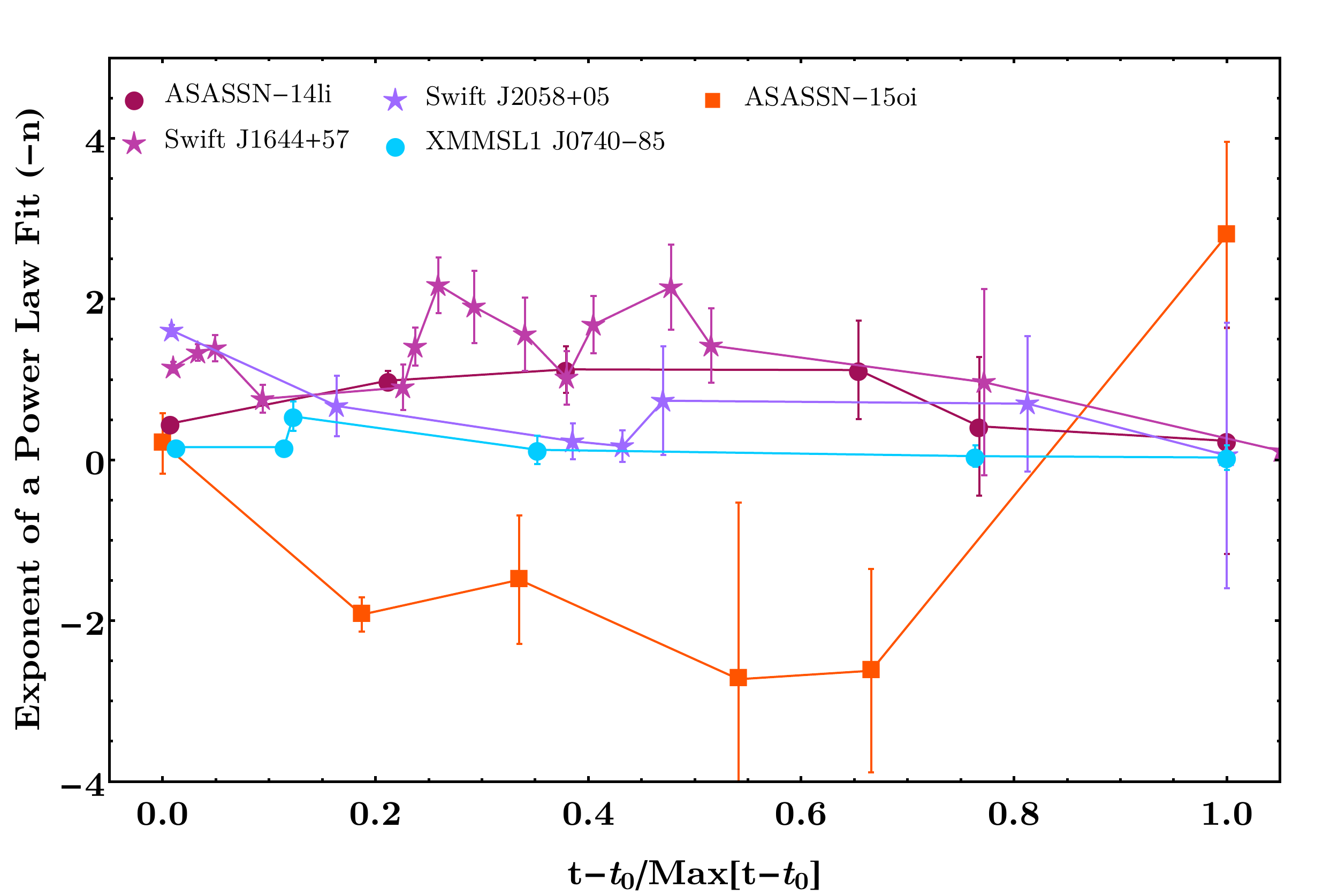}
		\caption{\textit{Upper:} Histogram of powerlaw indexes as seen at early (solid) and late (dashed) times during the decay of the X-ray emission from a sample of X-ray TDE candidates. Overlaid as the vertical solid, large dashed, dotted and dash-dotted grey lines are the powerlaw indexes for fallback (-5/3), advective, super-Eddington slim disk accretion (-4/3), viscous disk accretion (-19/16) and disk emission (-5/12).  Figure taken from \citet{Auchettl:2017a}. \textit{Lower:} Best fit powerlaw index and its uncertainty for the X-ray TDE sample of \citet{Auchettl:2017a}. \label{indexes}}
	\end{center}
\end{figure}

The evolution of a TDE's light curve depends heavily on a number of factors such as the stellar structure \citep[e.g.,][]{Guillochon13}, 
and whether the emission arises from fall-back \citep[e.g.,][]{Rees:1988a, 1989ApJ...346L..13E, Phinney:1989a}, disk emission \citep[e.g.,][]{Rees:1988a,Cannizzo:1990ar,Lodato:2011a} or super-Eddington accretion \citep[e.g.,][]{2011ApJ...742...32C}.
X-ray light curves of TDEs are rarely well enough sampled to be able to identify the time of disruption, or even the time of peak flux. Commonly, when fitting the light curve, the two parameters, $t_{0}$ and the index $\alpha$ are found to be degenerate and $\alpha$ is fixed to the canonical value of -5/3 to obtain an estimate of the date of disruption. 
 However,  studies of the optical emission from individual sources \citep[e.g.][]{Wevers:2019b}, and detailed global studies in the Far-UV \citep{vanvelzen:2019a} have shown that this behaviour is not necessarily universal, especially at late-times.   \citet{Auchettl:2017a}, took a different approach. Under the assumption that $t_{0}$ is the time of the first, peak flux, detection, they showed that the X-ray emission from TDEs then seems to decay with a power-law index that is shallower than $t^{-5/3}$, implying that the viscous timescale\footnote{Here the viscous timescale is defined as the time it takes for material to accrete onto a black hole, and depends on the height and radius of the disk and the orbital period \citep{2015ApJ...809..166G}.} is long for these events \citep{Guillochon13}. 
 At early times, TDEs 
 were found to have power-law indexes that are consistent with both fall-back ($t ^{-5/3}$) and disk emission ($t ^{-5/12}$)
 \citep[see][]{Lodato:2011a}, while at late times most, if not all, events tend to have decay rates consistent with disk emission (see Fig.~\ref{indexes} upper panel). The transition between these different emission processes is not necessarily smooth, with small timescale variations in the power-law index seen as each event evolves (Fig.~\ref{indexes} lower panel), while the time it takes for each event to undergo this transition varies between sources. Within the class of X-ray emitting TDEs, those whose emission is dominated by a strong jet (e.g., jetted TDEs such as Swift J1644+57: \citealt{2011Sci...333..203B, 2011Sci...333..199L,Burrows:2011a, 2016ApJ...819...51L}) showed multiple transitions between different emission processes, while those whose emission is not dominated by a jet (thermal TDEs such as ASASSN-14li: \citealt{2015Natur.526..542M, 2016MNRAS.455.2918H, 2017MNRAS.466.4904B, 2018MNRAS.475.4011B}) tend to show less variation in their power-law index which tends to fluctuate around the index associated with one type of emission process. 

Even though the X-ray light curves of TDEs tend to decay following a powerlaw index that is consistent with fall back, or disk emission, a handful of TDEs show strong deviations from this behaviour. Even though the light curve of the jetted TDE Swift J1644+57 globally exhibits an approximately $t^{-5/3}$ decay (which is punctuated by flaring, variability and dips) \citep{2011Sci...333..203B, 2011Sci...333..199L,Burrows:2011a,2016ApJ...819...51L, 2016ApJ...817..103M}, a striking feature of its light curve is the dramatic decrease in X-ray flux $\sim$500 days after its initial detection. Within $\sim$ 4 days, the X-ray flux dropped by a factor of $>50$, which corresponds to a decay index steeper than $t^{-70}$ \citep{2016ApJ...819...51L}. This behaviour suggests that the accretion underwent a state change; either it suddenly stopped, possibly consistent with the star being completely accreted onto the black hole \citep{2012MNRAS.419L...1Q}, or the accretion became sub-Eddington and radiatively efficient which dramatically represses power-law jets \citep[e.g.,][]{van-Velzen:2011b,2011ApJ...739L..19R,Zauderer:2013a}. 

\section{Indirectly identified X-ray events from reprocessing into high-ionization emission lines, and X-ray follow-up}

Some luminous X-ray TDE candidates were not detected directly in the X-ray regime, but the flaring X-ray emission was indirectly inferred based on the presence of luminous, then fading, high-ionization iron coronal emission lines
 in optical spectra. These lines need a strong incident X-ray continuum in order to be created \citep{Komossa:2008ar,Komossa:2009a,Wang:2011a,Wang:2012ar}. Only one of these events, 
 SDSS~J095209.56+214313.3, had X-ray follow-up spectroscopy which revealed a relatively flat X-ray spectrum,  and a strong decline in X-ray luminosity with $L_{\rm x, low} = 4 \times 10^{40}$ erg/s between 2--10 keV (Komossa et al. 2009). Selected non-X-ray properties of these events will be further discussed in the \echochap{}.

\section{Long-lived events}

\begin{figure}[]
  \centering
  \includegraphics[width=0.6\textwidth]{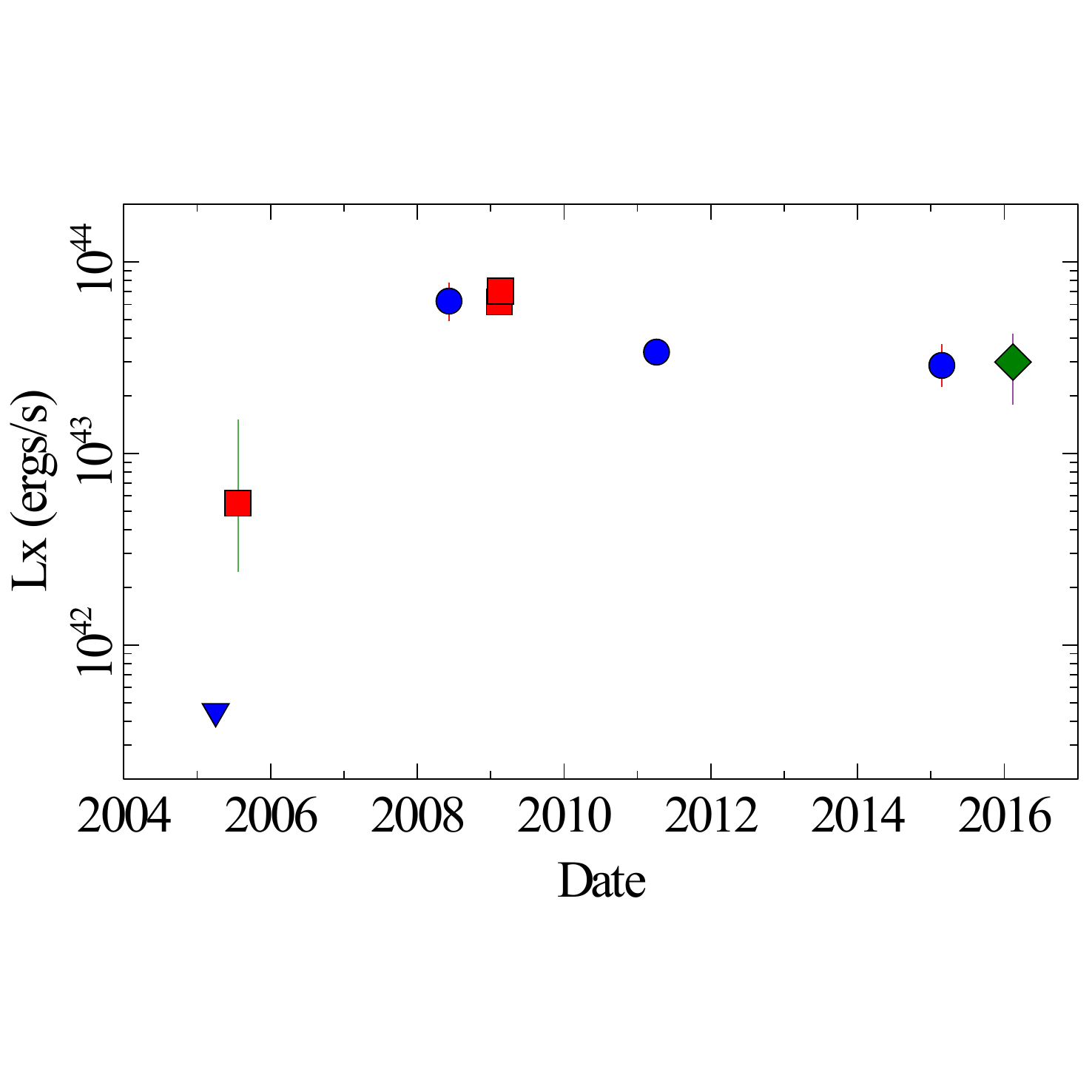}
  \caption{X-ray light curve of 3XMM~J150052.0+015452 from Chandra (blue circles and upper limit), XMM-Newton (red squares) and Swift-XRT (green diamond). Data taken from \citet{Lin:2017a}.}
  \label{fig:x:lc_1500}
\end{figure}

We have seen in the previous sections that some TDEs (e.g. NGC~5905, \msevenns, \sdsstwlong and IGR~J12580+0134) decay quickly from their
peak X-ray luminosity. Other events, however, maintain their peak emission for longer; NGC~3599 had a plateau of at least 18 months before decaying \citep{Saxton:2015a} and \mfourteen at least 100 days. The champion though is 3XMM~J150052.0+015452 \citep{Lin:2017a}, a TDE from a dwarf galaxy, which rose within 4 months in 2005 and has been decaying very slowly for more than ten years (Fig.~\ref{fig:x:lc_1500}). This event has maintained a soft spectrum, being well modelled with a low-temperature, kT$\sim40$ eV, black-body, heavily Comptonised by optically-thick, low-energy electrons ({\tt COMPTT} model), for the whole of its plateau phase.
This model, in stellar-mass black holes, is believed to signify a super-Eddington accretion state \citep[e.g.][]{Middleton:2013a}. The last Chandra observation, has a relative deficit of hard flux which can be interpreted as either; that the accretion mode has changed to a super-soft (sub-Eddington) state or that an ionised absorber was in the line-of-sight which
caused the sharp spectral drop towards higher energies.
The black hole mass is estimated to be $\sim10^{6}$ \msolar based on the mass of the host dwarf galaxy, consistent with the observed $L_{\rm bol}\sim10^{44}$ \lumunitsns. \citet{Lin:2017a} interpreted the slow evolution of the bolometric light curve as indicating a distant circularisation of the stellar debris, leading to higher viscosity and a consequently slow fallback of the material to the BH. The mass of the disrupted object is then around 2\msolar with $\sim0.9$ \msolar being converted into radiation.   
Two unusual persistent AGN have shown similar long high-luminosity, soft X-ray emission; GSN~069 \citep{Miniutti:2013a} and 2XMM~J123103.2+110648 \citep{Terashima:2012a}. Both of these have been interpreted by some authors as a TDE occurring in an AGN \citep{Lin_1231, Shu2018a}. 

There are several theoretical reasons for why a TDE may be long-lived. For example, a long super-Eddington accretion phase, perhaps involving a disrupted object with a large mass \citep[e.g.][]{Lin:2017a}; a partially stripped evolved star atmosphere \citep{MacLeod:2012ar} or late, distant circularisation \citep{2015ApJ...809..166G, Shiokawa:2015a}.

\section{Very fast events}

Some TDEs are expected to lead to fast X--ray flares. For instance, a TDE involving an intermediate-mass black hole (IMBH), defined here as a black hole with a mass less than 10$^5$ M$_\odot$ (e.g.~\citealt{Evans:2015a}). Especially if the disrupted star was a compact star, such as a white dwarf (see the \wdchap{}), then the associated orbital time scales are short and the relativistic periastron precession is large, potentially leading to a short circularization time (see the \disrupchap{}) and, therefore, a short rise time for the accretion flare (e.g.~\citealt{Clausen:2011ar, MacLeod:2016a,2018arXiv180909359S}). Alternatively, it is predicted that in some cases a shock occurs  the star upon disruption which breaks out of the star and gives rise to a brief X--ray flare (e.g.~\citealt{Yalinewich:2018ar}). Finally, blazar--like variability in those TDEs where a relativistic jet launched from the near--vicinity of the black hole is pointing close to our line of sight can give rise to fast X--ray variability even for black holes more massive than IMBHs (e.g.~\citealt{2011Sci...333..203B, 2011Sci...333..199L}).

Below we discuss the fast X--ray events that have been suggested to be caused by tidal disruption events, we do note, however, that relatively few things are known about these events, making their interpretation as being caused by a tidal disruption event much less secure than some of the other events of longer duration in this manuscript.

First of all, we need to be aware that several aspects can make an event appear ``fast''. For instance, if given the limited sensitivity of any instrument only the peak of a (longer duration) outburst is detected an event can appear as ``short'' or ``fast''. This effect probably affected earlier detections of fast events (e.g.~\citealt{1983MNRAS.205..875P}; \citealt{Grindlay1999}) more than recent detections with more sensitive instruments, although of course events at large(r) distances will still cause this effect even in modern detectors (as all detectors have a sensitivity limit). Furthermore, several minute--to--hour scale X-ray flares are known that have nothing to do with tidal disruption events, such as M--star flares (\citealt{1975ApJ...202L..73H}) and more generally, flares due to stellar activity often induced by binary interactions such as those of RS CVn stars (\citealt{1983MNRAS.205..875P}). And finally, accretion flares from Galactic low--and high--mass X--ray binaries  (\citealt{2000A&AS..147...25L}; \citealt{2001A&A...368.1021L}) and thermonuclear explosions on the surface of an accreting neutron star (so called Type I X-ray bursts) can appear as fast X-ray flares (e.g.~\citealt{2008ApJS..179..360G}). 

In order to weed out flares from such events multi--wavelength data is crucial. For that a source localization accurate to the order of arcseconds is important: this astrometric accuracy comes naturally with {\it Chandra}, XMM--{\it Newton} and {\it Swift}--XRT--discovered transients.

\begin{figure*}
\begin{center}
\includegraphics[width=1.0\textwidth]{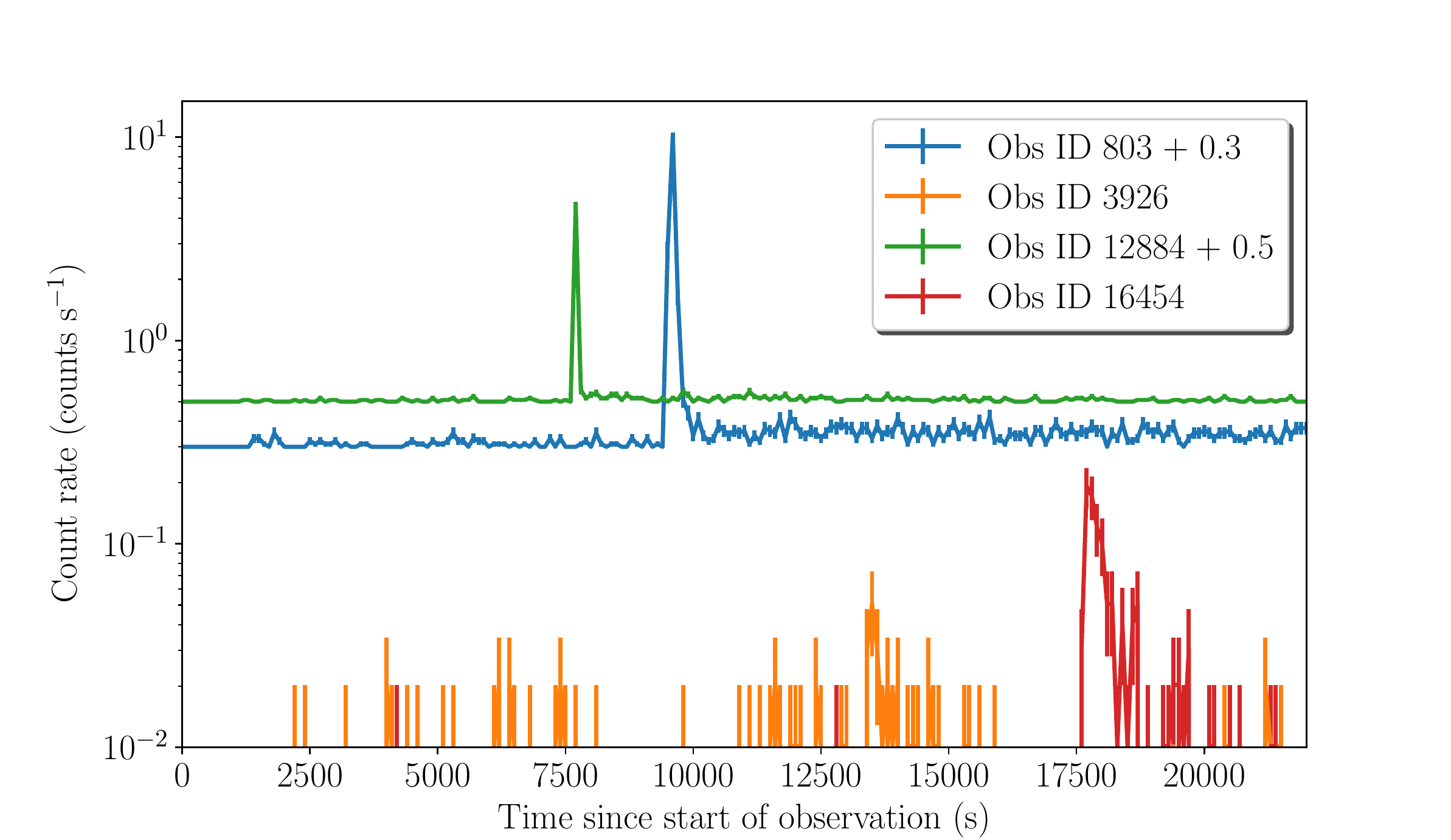}
\end{center}
\caption{\footnotesize
The light curves of the four fast X-ray transients discovered in {\it Chandra} data. The data has been binned in time intervals of 100 seconds. The time zero on the X-axis corresponds to the start of the observation in each case. The observation ID of the {\it Chandra} data is indicated in the top right side of the figure. Clearly, the detected count rate varies significantly between the four events. Furthermore, the duration of the flares differs as well. The flare detected in observation ID 12884 is the shortest.
}
\label{fig:lc-fast}
\end{figure*}

{\it XRT~000519:} The first of this new type of fast
transient X-ray sources (XRT~000519) was found in an archival \chan\,observation
(Observation ID 803; \citealt{2013ApJ...779...14J}; see Fig.~\ref{fig:lc-fast}). The source position lies 12.16
arcminutes from the centre of the Virgo Cluster galaxy M~86, but it
does not fall in the M~86 $\mu_B=25$ magnitude per arcsec$^2$ isophote
area. Optical Isaac Newton Telescope images show a tidal stream stripped off the galaxy
SDSS~J122541.29+130251.2, suggesting that M~86 is undergoing a minor
merger. The small projected distance on the sky between
the position of the transient and that of the tidal stream 
suggests that the transient is associated with M~86. Uniquely to this source (when compared to the others, see below) is that 
the main flare is double peaked. 

Deep observations with the William Herschel Telescope (WHT) r$^\prime\approx 25$ and $K\approx 20$--band observations rule out an
M-star flare and also globular cluster hosts brighter than M$_V\sim-6$ at
the location of M~86 for this event. 

{\it XRT~110203:} A second transient which has properties that strongly resemble those
of XRT~000519 is XRT~110203 ({\it Chandra} observation ID 12884; \citealt{2015MNRAS.450.3765G}). 
Also this event lies close to a cluster of galaxies (ACO~3581; z=0.023) but no host galaxy is detected so far.  
 
{\it CDFS-XT1:} \citet{2017MNRAS.467.4841B} report the discovery of a third fast X--ray transient in the
\chan\, Deep Field South survey (CDFS-XT1; {\it Chandra} observation ID 16454). The light curve indicates a
rise time of $\sim$70-160 s. In this case they found a possible host
dwarf galaxy in the 3D-HST field (with AB magnitude R$\sim$27.5), but
there is no cluster of galaxies nearby.  Unfortunately, there is no
spectroscopic redshift of this source.
The photometric redshift of $\sim$2 assigned to this host is
much larger than the redshifts associated with the first two
events. However, the redshift estimation is still very uncertain as
the host was only detected in two HST filters (ACS/F606W $\sim$V-band
and WFC3/F125W $\sim$J-band). 
In addition, the host--event association
also needs to be confirmed as the chance alignment probability for
such deep images is non-negligible and finally, the astrometric
alignment between the host and the transient event is not perfect.
 
Given that these three events share the same timescale, have no clear or
only a faint host, using Occam's razor, we suggest
that the three events are drawn from the same parent
population. Glennie et al.~(2015) estimate a rate of $\sim$10$^{5}$ events per year over the
whole sky with a peak X-ray flux greater than 10$^{-10}$ erg cm$^{-2}$
s$^{-1}$. 

\citet{2016Natur.538..356I} found one source to flare to a peak $L_{X}$ of $9\times10^{40}$\lumunits and 5 repeat flares to $L_{X}\approx10^{40}$\lumunits and the probable detection of persistent/long term X--ray emission from 
optical sources consistent with being a globular 
cluster or ultra--compact dwarf host. These flares have a similar time scale and
peak luminosities L$_X>10^{39}$ erg s$^{-1}$. Repeat flares have not been observed for the 
three X-ray flares XRT~000519, XRT~110203 and CDFS-XT1 discussed above, although 
the repetition time scale is $>$4  days for one and $\sim$1.8 days for the second repeating source in 
\citet{2016Natur.538..356I} and given the 
sparseness of the X--ray observations of the fields of the three X--ray flares above 
one cannot rule out that all events repeat with timescales of days to weeks. Whereas this 
might argue against cataclysmic events, such as TDEs, models predicting multiple flares due to for instance partial
disruptions of a white dwarf on an eccentric bound orbit or binary black holes might well still 
be consistent with the observations (\citealt{2010MNRAS.409L..25Z}). However, we do note that 
the predicted periods would be of order of hundreds to perhaps thousands of seconds, not days for the bound white dwarf orbiting an IMBH. 

For the flares where {\it Chandra} detected sufficient number of counts for a spectral decomposition, it was found that the spectra of the flares are well--fit with a power law shape (\citealt{2013ApJ...779...14J}; \citealt{2015MNRAS.450.3765G}), although in the latter case the value of the power law index depends strongly on the assumed interstellar extinction present. In the case of the flare called ``source 1'' in \citet{2016Natur.538..356I}, the spectrum was also well--fit by a power law with index 1.6$\pm$0.3 (90\% confidence), fixing the Galactic column density to the value of 1.8$\times 10^{20}$ cm$^{-2}$. The power law index varied slightly between the two peaks in the case of the XRT~000519 flare, with the second peak having a slightly softer power law index (1.95$\pm$0.05; 68\% confidence) than the first peak (1.6$\pm$0.1; 68\% confidence). A power law index of 1.4$\pm$0.2 (68\% confidence) provided a good fit for for the flare reported in \citet{2017MNRAS.467.4841B}, although those authors also warn that a softer index (closer to 2) can not be ruled out as the power law index and (Galactic) extinction are to a large degree degenerate given the relatively low number of counts detected.

Other fast X--ray flares, often with time scales of thousands of seconds, hence slightly longer 
than the $\sim$100 s time scale for the main flares above, have been reported as ultra-long gamma-ray bursts (\citealt{Levan:2014a}), although the energy bands used to measure the duration of the flares is different, making their comparison more difficult.

A particularly interesting case is CDFS-XT2 \citep{Xue:2019a}. In that paper it was interpreted as
the X-ray signal from a binary neutron star merger. However, \citep{Peng:2019a} interpret it as a white dwarf TDE. We note that the light curve \citep{Xue:2019a} is more like
the ultra-long GRBs \citep{Levan:2014a}, than the faster-still X-ray transients we report on here.
However, clearly there is overlap in the properties of these events.

The energies and timescales associated with these events imply that
compact objects such as massive black holes must be involved. Given
the expected rate of 10$^5$ over the whole sky per year, a conservative assumed $\log$
N - $\log$ S (similar to other observed X--ray sources where per 2
decades in luminosity ten times more sources are found [e.g.~\citealt{2012MNRAS.419.2095M}]), 
and the instantaneous sensitivity of the eROSITA satellite should find $\sim$1 of these fast X--ray events per day.

\section{Multi-waveband properties of X-ray selected TDEs}

\subsection{UV and optical}
\label{sec:x:multilambda:uv}

Among the first X-ray TDEs identified, NGC 5905 had quasi-simultaneous
optical photometry, thanks to photographic plate archives which covered a
timespan of several decades (Fig. 2 of Komossa \& Bade 1999). No long-term
optical variability of this HII-type galaxy was discovered, and no optical
flaring quasi-simultaneous with the X-ray flare was detected.

\begin{figure*}
\begin{center}
\rotatebox{90}{\includegraphics[width=0.6\textwidth]{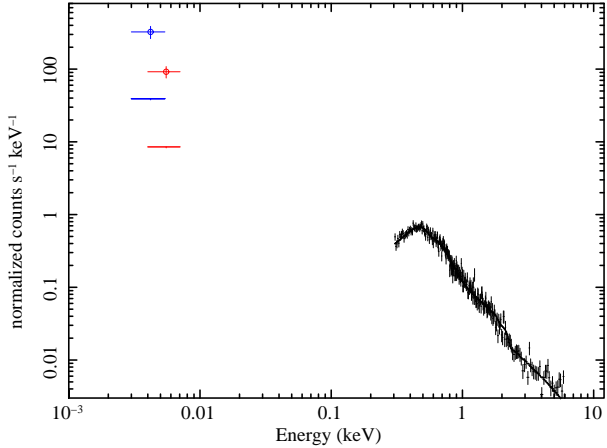}}
\end{center}
\caption{\footnotesize
An extrapolation of the best fit X-ray model, a power-law plus a kT=86 eV black-body, to the XMM-OM, UVW1 and UVM2 filter data  of \msevenns.
The single temperature thermal model significantly underpredicts the observed UV flux.
}
\label{fig:fullspec_0740}
\end{figure*}

The Swift and \xmm satellites both host an UV/optical telescope in their payload which allows these bands to be monitored simultaneously with the X-ray emission. 
Initially, it was thought that any emission in these bands would be from the Rayleigh-Jeans tail of the hot plasma 
which produces the soft X-ray emission and would therefore rise and decay simultaneously with the X-rays \citep[e.g.][]{1999ApJ...514..180U}.  
In \mseven the
UV and X-ray flux did decay from peak quasi-simultaneously
over 550 days \citep[Fig.~\ref{fig:x:lc_0740_1446}; ][]{Saxton:2017a}. Nevertheless, the galaxy-subtracted UV emission sits well in excess of a simple extrapolation of the thermal component from the X-ray spectrum of this source to UV wavebands (Fig.~\ref{fig:fullspec_0740}). 
The UV and X-ray data
can be well modelled by a structure which includes emission from a range of 
temperatures, such as a thin accretion disk \citep{Saxton:2017a}. This finding agrees with the low temperature emission of
$\sim20,000$K ubiquitously found in TDEs discovered in the optical or UV bands \citep[][and see the \optchap{}]{Gezari:2009ar, vanVelzen:2011ar, 2016MNRAS.463.3813H}. 
The similar temporal behaviour of the X-ray and UV emission is mirrored in ASASSN-14li \citep[section~\ref{sec:asassn-14li} and][]{2016MNRAS.455.2918H, 2017MNRAS.466.4904B} and indicates prompt accretion (or efficient circulation: \citealt{2015ApJ...809..166G}). In this event, the high-density, multi-wavelength monitoring indicated a possible delay of 32 days between the X-ray emission and that of the UV \citep[][and see the \echochap{}]{Pasham:2017a}.
If the UV is produced by shocks in colliding streams of debris \citep{Piran:2015b, Shiokawa:2015a} then it would naturally precede the X-ray emission which is generated when that same material falls down to the black hole. The timescale for the infall is indeed expected to be a few weeks \citep[][and the \disrupchap{} and \diskchap{}] {Piran:2015b,Shiokawa:2015a,Bonnerot:2017a}.

In ASSASN-15oi the behaviour was quite different (Sect.~\ref{sec:asassn-15oi}). Here the UV light fell by 5 magnitudes  over 200 days while the soft X-ray luminosity increased by a factor 10. The X-rays subsequently declined back to their initial level after a further 400 days.
\citet{2017ApJ...851L..47G} and \citet{2018MNRAS.480.5689H} presented late-time observations of ASASSN-15oi. Their studies revealed that the thermal X-ray emission from the source brightened by an order of magnitude during its first year of evolution, rather than following a powerlaw decline as seen from its optical/UV light curves. After $\sim600$ days, the X-ray emission had faded back to the levels originally detected at peak. 
\citet{2017ApJ...851L..47G} suggested that the $\sim$1 year it takes in the rise to peak for the X-ray emission is the result of delayed accretion on a $10^{6}M_{\odot}$ black hole. This delayed accretion is a result of inefficient circularisation of the debris disk due to a delay in the time it takes for the material to accrete onto the black hole \citep{Lodato:2012a,Piran:2015b,2015ApJ...809..166G}. However, as the circularisation timescale is proportional to $M_{\rm BH}^{-7/6}$ \citep{Bonnerot:2017a}, \citet{2017ApJ...851L..47G} also suggested that if the black hole of ASASSN-15oi was closer to $10^{7}M_{\odot}$ as originally estimated by \citet{2016MNRAS.463.3813H} using the host galaxy mass and the $M_{\rm BH}-M_{\rm bulge}$ relation of \citet{2013ApJ...764..184M},  then the circularisation timescale would be much shorter, and may not be able to explain the observed behaviour.
Another possibility is that ASASSN-15oi is heavily absorbed by material surrounding the accretion disk which is optically thick to X-ray radiation at early times \citep[e.g.,][]{Metzger:2016b}. Even though, observationally  \citet{2016MNRAS.463.3813H, 2017ApJ...851L..47G} and \citet{2018MNRAS.480.5689H} were unable to constrain the column density in the direction of the source to confirm this suggestion, the possibility that some X-ray TDEs are surrounded by dense material may be  supported by the study of \citet{Auchettl:2017a}.

A still different picture is present in the light curve of \mfourteen (Fig.~\ref{fig:x:lc_0740_1446}). 
Here the UV flux from the galaxy increased by $\sim1$ magnitude, prior to the first X-ray detection and then stayed flat for 400 days before fading by $\sim$ a magnitude over the next 200 days.
Meanwhile, the X-ray emission was constant for 100 days after discovery before fading by a factor $\sim100$ over the next 500 days \citep{Saxton:2019a}. The delayed decay of the UV emission in this event suggests that a reservoir of relatively cool material was maintained for about a year. One possible explanation is that the material formed an accretion disk which drained viscously until running out of material \citep{vanvelzen:2019a}.

In \sdsstw the UV emission from the galaxy was apparently unaffected by the disruption event, while the X-rays
faded by a factor 100. While we should note that a constant UV/optical flux can simply mean that the flare is obscured in these bands or the contrast with the bright host galaxy is poor, the diversity of relative behaviour between the optical, UV and X-ray bands represents a challenge to current models of TDE evolution.
A full understanding of the diversity of optical/UV emission behaviour from X-ray selected TDEs awaits a better understanding of the  mechanisms which are responsible for the optical/UV emission. This theme is addressed in the \optchap{}.

\subsection{Radio}
Dedicated radio follow-up was performed on the TDE in NGC 5905, in order 
to exclude the (very unlikely) scenario of a blazar hiding in this nearby starburst galaxy, and 
to search for the first time for jet emission associated with a TDE itself \citep{Komossa:2002a}.
Based on an observation with the VLA A array carried out in 1996, no nuclear radio emission was detected, with a 5$\sigma$  upper limit for the presence of a central point source of $<150$ $\mu$Jy \citep{KomossaDahlem:2001a,Komossa:2002a}. At a distance of 75.4 Mpc of NGC 5905, this corresponds to a limit of 
$L_{\rm 8.46 GHz} < 9\times10^{36}$ \lumunitsns. Extended low-level radio emission at lower frequencies is present in NGC 5905, and consistent with its HII-type nature (see Sect. 2.3.2 of Komossa 2002).  A search for late-time radio emission from the nucleus of NGC 5905 was carried out by \citet{Bower:2013a}, who did not detect any to an upper limit of 200 $\mu$Jy. Radio emission from RXJ~1242-1119 was searched for in the FIRST VLA sky survey at 20cm \citep{Komossa:2002a} but none was detected. In order to search for very late-time radio emission, \citet{Bower:2013a} also carried out follow-ups of the other ROSAT TDEs. No radio emission was found from RXJ~1242-1119, while a second source in the error circle of RXJ~1420.4+5334 does emit faintly ($114\pm{24}$ $\mu$Jy) in the radio regime. 

Following the Swift detection of a jetted TDE, radio follow-up was more routinely carried out for newly identified TDEs. Radio upper limits of 100-200 $\mu$Jy were reported for \sdsstw \citep{Saxton:2012a} and 10 $\mu$Jy for \mfourteen \citep{Saxton:2019a}. Faint radio emission (1.2 mJy at 1.5 GHz; $L_{\rm 1.5 GHz}=10^{37}$ \lumunitsns) was detected 1 year after discovery from \msevenns, which faded over the  following months \citep{Alexander:2017a}, while slightly stronger emission was monitored in ASASSN-14li \citep{Alexander:2016a}. 

In summary, the thermal X-ray TDEs are not strong radio emitters. For a full analysis of the radio properties of
these and other TDEs see the \radiochap{}.

\section{Interpreting X-ray TDEs with a reprocessing model}
\label{sec:xreproc}
Various authors have explored the effect on the emitted radiation of an optically-thick, static or expanding envelope of material created during the event \citep{1999ApJ...514..180U,Strubbe:2009ar,Metzger:2016b,2018ApJ...859L..20D,Roth:2018a}. In this section we compare this model with X-ray TDE properties.
The X-ray emission from a (non-jetted) TDE is generally quite soft in nature, and is well described by a black-body with a temperature between 10-100 eV (see the previous sections)
or a powerlaw with photon index of $\Gamma>4$ \citep[][and earlier work]{Auchettl:2017a}. 
These temperatures are generally consistent with the picture suggested by \citet{Rees:1988a} of a black-body with a temperature between $10^{5-6}$ Kelvin arising from an accretion disk \citep[e.g.,][]{1999ApJ...514..180U, 2007ApJ...659..211B}. However, the temperatures derived from optical/UV studies of these events and optical/UV only events are an order of magnitude less than those measured from X-rays (see the \optchap{}) and may feasibly represent emission from the reprocessing of nuclear radiation  \citep[e.g.][]{Loeb:1997a}. 

Those that have a strong jet such as Swift J1644+57 and Swift J2058+05 \citep{Cenko:2012b} generally exhibit much harder X-ray emission, which can be best described by a simple powerlaw with a photon index of $\Gamma \sim 1-2$. There are also cases of non-jetted TDE, such as ASASSN-15oi, 
XMMSL1 J0740-85 and PS18kh \citep{2018arXiv180802890H,vanVelzen:2019b} which exhibit both the soft black-body component and a weaker powerlaw component with a temperature and photon index similar to that listed above (see Sect.~\ref{sec:x:spectral_properties}).
This diversity in the observational characteristics of these events has been suggested to be a natural result of the viewing-angle with respect to the orientation of the accretion disk \citep[e.g.,][]{2018ApJ...859L..20D}, or that X-ray TDEs and optical/UV only TDEs could result from a separate class of events that have compact debris disk due to large apsidal precession of the self-intersecting streams of the disrupted star \citep[e.g.,][]{2015ApJ...812L..39D}. 
\citet{Jonker:2019a} discuss three optically-selected TDEs which were not detected in X-rays during the optical flare but 8-10 years later had luminosities of $L_{X}\sim10^{41-42}$. In the reprocessing model, the X-ray emission from these events will have been massively suppressed at early times and only become visible when the density of the absorbing material decreased.

\begin{figure}
	\begin{center}
		\includegraphics[width=0.7\textwidth]{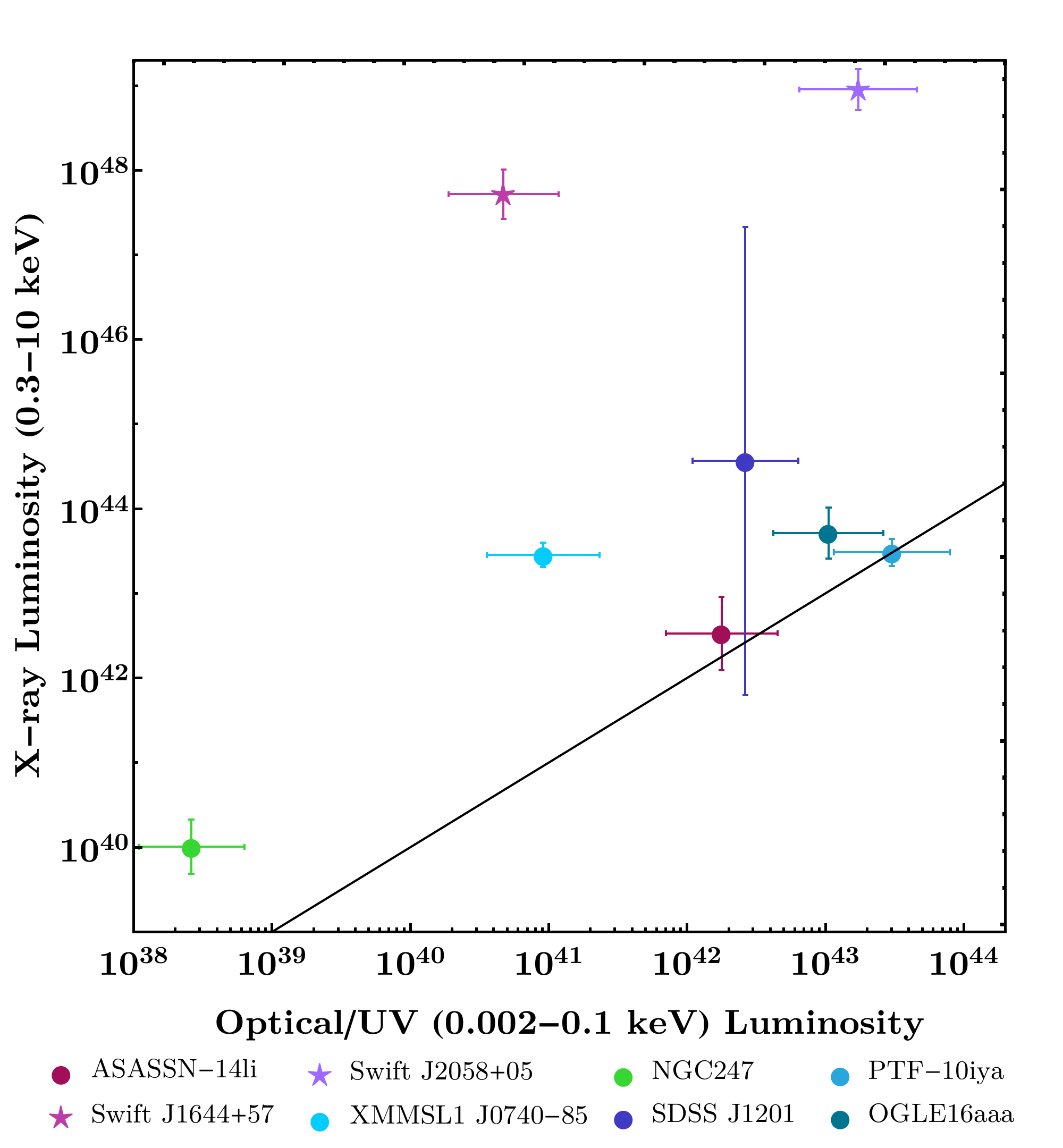}
		\includegraphics[width=0.9\textwidth]{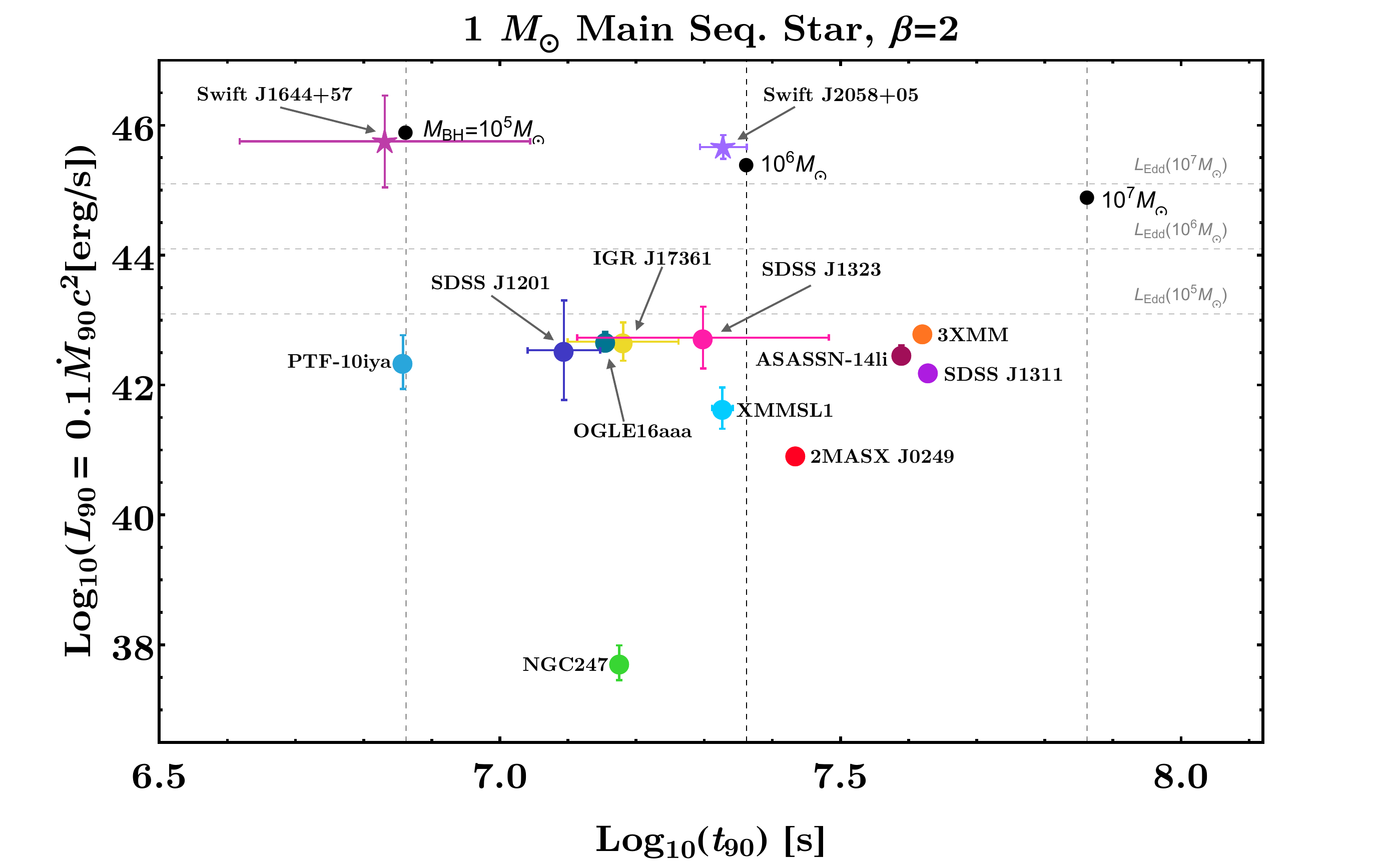}
		\caption{Upper: Peak X-ray v peak optical/UV luminosity for a selection of X-ray bright TDEs. Lower: The mean isotropic luminosity plotted against the time taken for the TDE to emit between 5 and 90\% of its total output. Both plots are taken from \citet{Auchettl:2017a}. 
		\label{xlumolum_t90s}}
	\end{center}
\end{figure}

Observationally, it has been shown that the softness of non-jetted events is intrinsic to the source, with the spectral energy distribution of these events peaking in the UV/soft X-ray band \citep[][and references in Sect. \ref{sec:rosat},\ref{sec:x:xmm_chandra}]{Auchettl:2017a}, consistent with theoretical expectations for black holes with masses $<10^{7}M_{\odot}$ \citep[e.g.,][]{1999ApJ...514..180U, 2018ApJ...859L..20D}. Using the spectral energy distributions of these events, \citet{Auchettl:2017a} saw that X-ray selected TDEs have high X-ray to optical ratios (see Fig.~\ref{xlumolum_t90s}). The fact that these events produce significant amounts of both X-ray and optical/UV emission opens the possibility that a considerable fraction of their emission is being reprocessed into optical/UV wavelengths \footnote{An alternative explanation was given by \citet{vanvelzen:2019a} who suggested that the properties of these sources are not a result of reprocessing but are due to a viscously spreading, unobscured accretion disk. This work was extended into the X-ray regime in \citet{Jonker:2019a}, who infer the existence of
a long-lived accretion disk to explain the relatively high late-time X-ray luminosity of
three optically-selected TDEs.}, with the variation seen in Fig.~\ref{xlumolum_t90s} perhaps implying that these events experience significantly different reprocessing rates. 

A separation between jetted and non-jetted events is seen when one attempts to derive the isotropic luminosity of each source 
\citep[][ Fig.~\ref{xlumolum_t90s}]{Auchettl:2017a}. Even though TDEs exhibit a wide range of isotropic luminosities, jetted events tend to have $L_{\rm iso}\sim10^{44}$ erg s$^{-1}$, while non-jetted X-ray events 
have $L_{\rm iso}\lesssim10^{42}$ erg s$^{-1}$
\footnote{Here $L_{\rm iso}$ is defined as the mean isotropic luminosity, after correcting for beaming, emitted by the event in the interval where the light curve contains between 5\% and 95\% of the total emitted luminosity (see \citet{Auchettl:2017a} and references therein 
for an explanation of the derivation of isotropic luminosity.)}.
TDEs which have been detected in optical/UV wavelengths only, known as ``veiled X-ray TDEs'' in the nomenclature of \cite{Auchettl:2017a}, could be surrounded by very dense material \citep[e.g., PS1-10jh:][]{2014ApJ...783...23G}, which would lead to the X-rays being reprocessed completely into optical/UV wavelengths \citep[e.g.,][]{2018ApJ...859L..20D}.  These optical/UV sources have isotropic luminosities that fall between the jetted and non-jetted X-ray TDEs, in what \citet{Auchettl:2017a} refer to as a ``reprocessing valley'' (see e.g., \citealt{2012Natur.485..217G, 2014ApJ...780...44C, 2014ApJ...783...23G, 2015ApJ...815L...5G, Piran:2015br}). As such, these events could emit X-rays but have all their emission reprocessed into lower wavelengths.

Assuming that a main sequence star is being fully disrupted, \citet{Auchettl:2017a} find that the light curves of X-ray TDEs are consistent with a disruption from a black hole with mass between $10^{5-7}M_{\odot}$. These results are consistent with that derived from e.g., modelling the light curves of the individual events \citep[e.g.,][]{2018arXiv180108221M}, from bulge-black hole mass relations of e.g., \citet{2013ApJ...764..184M, 2015ApJ...813...82R}, and optical spectroscopy \citep{2017MNRAS.471.1694W}.  

\section{TDEs in binary SMBHs and recoiling SMBHs}
\label{sec:x:binarySMBH}

\subsection{TDEs in binary SMBHs}
The appearance of TDEs which occur in binary SMBHs can be different from TDEs  of single SMBHs, including lightcurves which look characteristically different \citep{Liu:2009a,Coughlin:2017a}, and including rates which can be boosted by up to several orders of magnitude in some stages of binary evolution \citep[e.g.][]{Chen:2009ar}.

The lightcurve of the TDE from \sdsstwlong (Sect.~\ref{sec:sdss1201}; Fig.~\ref{fig:x:lc_1201}) does not show a smooth decline, but exhibits episodes of dipping.  One month after the peak, the X-ray emission suddenly dropped by a factor of $>50$ within a week and the source was no longer detected by Swift. X-rays re-appeared after 115 d, and then dropped a second time.
While such a behaviour could arise due to beaming in jetted sources, no radio emission was detected from this TDE \citep{Saxton:2012a}.
Instead, the characteristic intermittence and recovery of the lightcurve of \sdsstw is reminiscent of predictions by \citet{Liu:2009a}, who computed TDE lightcurves in binary SMBHs. In that case, the second SMBH acts as a perturber and the accretion stream on the primary is temporarily interrupted. Simulations by \citet{Liu:2014a} have shown, that the lightcurve of \sdsstw is consistent with a binary SMBH model with a primary mass of 10$^6$ M$_\odot$, a mass ratio $q \sim 0.1$ and a semi-major axis of 0.6 milli-pc.

This was the first supermassive binary BH (SMBBH) candidate identified in a non-active host galaxy, and the one with the most compact orbit among the known SMBBH candidates \cite[review by][]{KomossaZensus:2016a}. It has overcome the so-called ``final parsec problem'' \citep[e.g.][]{Colpi:2014a}. 
Upon coalescence, it will be a strong source of gravitational wave emission in the sensitivity regime of the upcoming generation space-based gravitational wave detectors.
If significant numbers of SMBBHs exist at the cores of non-active galaxies, we expect to see more such events in well-sampled lightcurves of TDEs with \swift or the future Einstein Probe \citep[EP;][]{Yuan:2015a,Yuan:2016a} mission. A good lightcurve coverage is essential for constraining the system parameters.

\subsection{TDEs as signposts of recoiling SMBHs}

Luminous X-ray flares from TDEs which occur {\em off-nuclear} are possible signposts of recoiling SMBHs \citep{KomossaMerritt:2008ar}. At X-ray peak luminosities in the quasar regime ($L_{\rm X} \sim 10^{42-46}$ erg/s), there is no other mechanism, which could produce a long-lived off-nuclear X-ray flare. 

Potentially, a flaring recoiling SMBH could hide among the population of ultraluminous X-ray sources (ULX) which have been identified in nearby galaxies, however, the known ULXs have much lower X-ray luminosities than AGN, and they are by far (i.e., by orders of magnitude) too abundant to be explained by TDEs from recoiling SMBHs \citep[see, e.g., the discussion by][]{Strateva:2009a, Jonker:2012a}.

\section{X-ray TDE rates}
\label{sec:x:rates}

To calculate the frequency of tidal disruptions it is necessary to perform a uniform analysis of a well-controlled sample of events.
\citet{Donley:2002ar}, performed
a systematic survey of galaxies observed in both the \rosat All-Sky Survey and
subsequent \rosat pointed observations, re-detecting 3 TDEs which
had been previously reported (see Sect.~\ref{sec:rosat}). From this
work, which covered 9\% of the sky, they calculated
a TDE rate of $9\times10^{-6}$ \galUnitsns. \citet{Esquej:2008a}
found 2 TDEs in the \xmm slew survey by comparing fluxes with
\rosat observations taken 10-15 years earlier and derived a rate of
$2.3\times10^{-4}$ \galUnitsns.
A further search for events was made by comparing RASS data with deep \xmm pointed observations \citep{Khab:2014_rosat} giving a baseline of 10--20 years. They discovered
three events in the 2\% of the sky covered by \xmmns, obtaining a rate of
$3\times10^{-5}$ \galUnitsns.
Finally, from multiple observations of the large cluster of galaxies, A1689, \citet{Maksym:2010ar} derived a
rate of $6\times10^{-5}$ TDEs \galUnitsns.

A survey measures three main quantities: the number of detected TDE,
the number of square degrees covered and the flux limit reached.
A set of assumptions are then adopted to find the volume of sky which has been sampled, and hence the number of galaxies observed, and the fraction of
a one-year light curve which has been observed. To convert these into a rate,
assumptions have to be made about the peak luminosity ($L_{P}$) of the event, the
galaxy density and the shape of the light curve (for a full description 
of the derivation see the \ratechap). In Table~\ref{tab:rates_surveys} we list the
measured values and adopted assumptions for each of the soft X-ray TDE rate calculations. We see that the assumptions vary greatly between each calculation.
$L_{P}$ has been taken to be between $2.8\times10^{43}$ \citep{Donley:2002ar} and
$\sim1\times10^{44}$ \citep{Khab:2014_rosat,Esquej:2008a} \lumUnitsns. Similarly the light curve shape has been
taken as effectively flat for one year \citep{Donley:2002ar},  flat
for 0.19 years and then dropping to zero \citep{Khab:2014_rosat} or $t^{-5/3}$ \citep{Esquej:2008a,Maksym:2010ar}. Integrating over the latter
gives a surveyed volume equivalent to observing the peak luminosity for 0.013 years. Not surprisingly these
different assumptions produce large differences in the final calculated rates,
which range from $9\times10^{-6}$ to $2.3\times10^{-4}$ \galUnits.

To show the importance of the details of the calculation, we take as a common set of assumptions, the peak luminosity from \citet{Donley:2002ar}, $L_{P}=2.8\times10^{43}$ \lumunitsns,
the visibility time from \citet{Khab:2014_rosat} (0.19 years) and a galaxy density of
$\rho=0.02$ Mpc$^{-3}$ and apply it to the surveys\footnote{Note that the cluster survey of \citep{Maksym:2010ar} uses a self-consistent set of assumptions which are not affected by this change.}. Results are shown in Table.~\ref{tab:rates_surveys} where the rates now
vary between 3.4 and 21$\times10^{-5}$ \galUnits, within a factor 7 of each other\footnote{All the surveys use very strong TDE candidates when calculating the rates except for \citet{Khab:2014_rosat} which identified one very likely TDE (RBS~1032) and two possible TDEs in their sample. If only 
RBS~1032 had been adopted here then the survey TDE rates with common assumptions would agree to a factor $\approx 3$.}. This immediately shows that the {\em measurements} in these surveys are actually in better agreement than they first appear.

For an accurate calculation of the absolute value of the TDE rate, a larger and less-biased sample of TDEs are needed to constrain their  properties. In particular the TDE peak luminosity function is not currently well constrained \citep[although see][for first attempts]{Sun:2015a,Auchettl:2018a} and the large variety of TDE light curves (e.g. Fig.~\ref{fig:x:lc_0740_1446}) introduce considerable
uncertainty in the rates which will only be resolved when an
unbiased estimate of the median light curve is available.

In summary, the {\em reported} soft-X-ray-selected TDE rates in the literature lie between $\sim 1\times10^{-5}$ and $2\times10^{-4}$ TDEs \galUnitsns. If we adopt a common set of assumptions, then the surveys agree to better than a factor 7.
Note that the measured rate is a lower limit due to the fraction of TDEs which are not observed in X-rays due to absorption by gas in the host galaxy, tidal debris or outflowing material from the accretion process, e.g. \citet{Sembay:1993a} estimate that 
40\% of soft X-ray TDEs will be located in edge-on galaxies and their soft X-rays absorbed away within the host galaxy. 

\section{Conclusions and future prospects}
\label{sec:x:conclusions}
The bulk of the TDEs discussed in Sect. 2--5 share the following characteristics:

\begin{itemize}
  \item X-ray peak luminosities between $10^{42}$ and a few times 10$^{44}$ \lumunits
  \item Very soft X-ray spectra near peak, with black-body temperatures in the range $kT_{\rm bb}$ = 0.04--0.12 keV (or, alternatively, with powerlaw indices in the range $\Gamma_{\rm x}=3-5$).
  \item A spectral hardening within years.
\item A decline on a timescale of years down to a quiescent level. 
\item An absence of X-ray emission lines
\item Fast variability in several events (minutes to hours). 
\item Host galaxies which are quiescent and in-active both before and after the disruption.
\item A decline in flux by factors up to 1000--6000. 
\end{itemize}

While there are certainly exceptions to these properties, the overall picture is of X-ray radiation emitted by optically-thick material within a few $R_{g}$ of the black hole, which fades with the diminishing return of tidal debris. 

The limited duration of events has allowed us to follow the passage from super-Eddington to sub-Eddington to low-level (RIAF/ADAF) accretion on year-to-decade timescales and witness the onset and cessation of disk winds which are responsible for driving material into the host galaxy.

One of the exciting prospects of X-ray TDE observations is the possibility of identifying IMBH from their distinctive disruptive properties. This was broached in Sect.\ref{sec:x:imbh} and also in the section on the enigmatic {\em very fast events}, whose light curves may indicate the disruption of a compact star by a BH of mass $\le10^{5}$ \msolar.  

Deep follow-up observations of TDEs near their peak are allowing us to probe the extremes of accretion physics in relatively clean environments.
They allow us to follow the evolution of
disk winds and coronae, search for relativistic (precession)
effects in the Kerr metric, 
estimate BH spin, carry out absorption/emission-line spectroscopy of ionized matter in outflow (either stellar
debris or accretion disk winds), and study the jet-disk coupling and jet evolution in jetted events.

Although more than 20 years old, the science of X-ray emitting TDEs can be considered to be still in its infancy, with each new event presenting traits which modify our understanding of the disruption process. With the launch of the eRosita telescope \citep{Predehl:2010a} on board the Spectrum-Roentgen-Gamma (SRG) mission, it is expected that hundreds of new TDEs will be found \citep{Khab:2014_erosita,Jonker:2019a}. If well monitored in dedicated follow-ups by other missions, 
these will fill in the parameter space and provide us with a more complete picture of the phenomenon. Within a few years, the Einstein Probe will provide excellent light curves of the rise, peak and initial decay phases of hundreds of X-ray TDEs and allow detailed modeling of the fall back process of a large number of TDEs.

\begin{acknowledgements}
Dacheng Lin and Erin Kara are warmly thanked for providing updated figures of their work. RS would like to thank Peter Maksym for early help with the
rates calculation. Sjoert Van Velzen and an anonymous second referee
are thanked for comments and suggestions which improved the manuscript.
\end{acknowledgements}


\clearpage
\begin{landscape}

\begin{table*}[ht]
{
\caption{Summary of soft X-ray TDEs referred to in this chapter.}
\label{tab:tde_summary}      
\begin{center}
\begin{small}
\begin{tabular}{lccccl}
\hline\hline                 %
Name & Date$^{a}$ & z & $L_{\rm X,peak}^{b}$ & Spectrum$^{c}$ & Ref$^{d}$ \\
\hline\noalign{\smallskip}
NGC 5905 & 7/1990 & 0.011 & $7\times10^{42}$ & $kT=60$ eV & \citep{Bade:1996a}\\
RXJ1624+7554  & 8/1990 & 0.064 & $2\times10^{44}$  & $\Gamma=3$ & \citep{Grupe:1999a}\\
RBS~1032 & 11/1990 & 0.026 & $1\times10^{43}$ & $kT=120$ eV & \citet{ghosh:2006a}$^{e}$ \\
RXJ1420+5334  & 12/1990 & 0.147 & $2.5\times10^{44}$ & $kT=38$eV & \citep{Greiner:2000ar}\\
RXJ~1242--1119 & 7/1992 & 0.05 & $4\times10^{44}$  & $kT=60$ eV & \citep{KomossaGreiner99}\\
TDXF1347-3254  & 12/1992 & 0.037 & $7\times10^{42}$ & $kT=120$ eV & \citep{Cappelluti:2009ar}\\
WINGS J1348  & 12/1999 & 0.063 & $1\times10^{42}$ & $kT=84$ eV & \citep{Maksym:2013ar}\\
3XMM~J152130.7+074916 & 8/2000 &  0.179 & $5\times10^{43}$ & $kT\sim170$ eV & \citep{Lin_1521}\\
NGC 3599 & 11/2003 & 0.002 &$5\times10^{41}$ & $kT=95$eV & \citep{Esquej:2007a}\\
SDSS J132341.97+482701.3 & 12/2003 & 0.0875 & $5\times10^{43}$ & - & \citep{Esquej:2007a}\\
SDSS~J131122.15-012345.6     & 2/2004 & 0.18 & $5\times10^{42}$ & kT$\sim100$ eV & \citet{Maksym:2010ar}\\
XMMSL1~J024916.6-041244  & 7/2004 & 0.0186 &$2\times10^{42}$ & $kT=110$ eV & \citep{Esquej:2007a}\\
3XMM~J150052.0+015452 & 7/2005 & 0.145 & $6\times10^{43}$ & kT$\sim40$ eV & \citep{Lin:2017a} \\
SDSS~J095209.56+214313.3 & 12/2005 & 0.079 & - & - & \citep{Komossa:2008ar}\\
3XMM~J215022.4-55108  & 5/2006 & 0.055 & $7\times10^{42}$ & $kT\sim280$ eV & \citep{Lin_2150}\\
2XMMi~J184725.1-631724 & 9/2006  & 0.0353 & $3\times10^{43}$ & kT$\sim80$ eV &  \citep{Lin_1847_2011}\\
\sdsstwlong  & 6/2010  & 0.146 & $3\times10^{44}$ & $kT=70$ eV & \citep{Saxton:2012a}\\
\mseven & 4/2014 & 0.0173 & $2\times10^{43}$ & kT$\sim80$ eV, $\Gamma_{\rm x} \sim 2$& \citep{Saxton:2017a}\\
ASASSN-14li & 11/2014 & 0.0206 & $3\times10^{43}$ & $kT=50$ eV & \citep{2016MNRAS.455.2918H}\\
ASASSN-15oi & 8/2015 & 0.0479 & $6\times10^{42}$ & $\sim$45 eV & \citet{2017ApJ...851L..47G}\\
\mfourteen  & 8/2016  & 0.029 & $6\times10^{42}$ & $\Gamma_{\rm x}\sim2.6$ & \citep{Saxton:2019a}\\
 
\noalign{\smallskip}\hline
\end{tabular}
\\
\end{small}
\end{center}
$^{a}$ Date of first detection in X-rays.\\
$^{b}$ The highest unabsorbed X-ray luminosity measured. Units of \lumunitsns.\\
$^{c}$ Dominant spectral component(s) at peak. "kT" refers to a black-body temperature,
$\Gamma$ refers to a power-law slope.\\
$^{d}$ The first reference to the X-ray measurements.\\
$^{e}$ Identified as a TDE by \citet{Maksym:2014ar} and \citet{Khab:2014_rosat}.\\
}
\end{table*}

\begin{table*}[ht]
\begin{center}
\caption{Summary of TDE survey rates}
\label{tab:rates_surveys}      
\begin{tabular}{l c l l l l l l l l l l l}
\hline\hline                 
\\

Survey $^{a}$ & Factor$^{b}$ & Lum$_{X}^{c}$ & Decay$^{d}$ & Baseline & $\rho_{\rm gal}^{e}$ & $F_{\rm lim}^{f}$ & $D_{\rm lim}^{g}$ & Nsrc$^{h}$ & Area$^{i}$
& Rate$^{j}$ & Rate$^{k}$ \\
         &           &  (erg/s)    & model & (Years)  & (Mpc$^{-3}$) & (\fluxunits)& (Mpc)     &  & (\% sky) & $gal^{-1}yr^{-1}$ & $gal^{-1}yr^{-1}$\\
\hline
RASSvRXP  & 20 & $2.8\times10^{43}$  & 1 yr & 0.5--8  & 0.014 &  $2.0\times10^{-12}$ & 342 & 3 & 9 & $9\times10^{-6}$ & $3.4\times10^{-5}$\\
RASSvXMMP & 10 & $10^{44^{l}}$ & 0.19 yrs & 10--20 & 0.02 & $3.5\times10^{-13}$ & 877 & 3 & 2 & $3\times10^{-5}$ & $2.1\times10^{-4}$ \\
XMMSvRASS  & 20 & $10^{44}$   &  $t^{-5/3}$ &  10--18 & 0.023 & $5.0\times10^{-12}$ & 406 & 2 & 15 & $2.3\times10^{-4}$ & $1.1\times10^{-4}$\\
Abell 1689 & 4--5 & $\sim10^{44}$ & $t^{-5/3}$ & 1--7 & - & $\sim1\times10^{-13}$ & - & 1 & - & $6\times10^{-5}$ & $6\times10^{-5}$\\

\hline                        
\end{tabular}
\hfill{}
\\
\end{center}
$^{a}$ RASS v \rosat pointed (0.2--2.4 keV) \citep{Donley:2002ar};
RASS bright source catalogue v \xmm pointed (0.2--2 keV) data \citep{Khab:2014_rosat}; 
XMM, EPIC-pn slew (0.2--2 keV) from catalogue XMMS1d3  v RASS \citep{Esquej:2008a} 
; repeat observations of 2100 galaxies in Abell 1689 over 7 years \citep{Maksym:2010ar}.\\
$^{b}$ Minimum variability factor needed for a detection.\\
$^{c}$ Assumed peak X-ray luminosity.\\
$^{d}$ Assumed light curve duration or decay function; flat for 
1 year \citep{Donley:2002ar}, flat for 0.19 years then dropping to zero
\citep{Khab:2014_rosat} or $t^{-5/3}$. \\
$^{e}$ Density of galaxies in the local universe.\\ 
$^{f}$ Detection flux limit (observed) in the 0.2-2 keV band. \citet{Khab:2014_rosat}
use the ROSAT Bright Source Catalogue which contains sources with
a minimum count rate of 0.05 c/s in the ROSAT PSPC camera. This translates to a 0.2--2 keV
flux of $3.5\times10^{-13}$ \fluxunits using their spectral model of a kT=50 eV black body absorbed by a 
Galactic column of $N_{H}=5\times10^{20}$cm$^{-2}$.\\
$^{g}$ Derived limiting distance for TDE detection.\\
$^{h}$ Number of TDEs found in the survey. Note that \citep{Donley:2002ar} consider sources
which have faded between the RASS and \rosat pointed observations and hence exclude RXJ~1242-1119.
\\
$^{i}$ Percentage of the sky covered in the survey. In the case of
Abell 1689, 2100 galaxies were repeatedly observed.\\
$^{j}$ The original reported TDE rate based on the given assumptions.\\
$^{k}$ The TDE rate calculated under common assumptions; $L_{\rm X,peak}=2.8\times10^{43}$,
an average visibility time of 0.19 years and a galaxy density of
$\rho=0.02$ Mpc$^{-3}$ (see text).\\
$^{l}$ \citet{Khab:2014_rosat} use a bolometric luminosity 
of $7\times10^{44}$\lumunits without explicitly giving the
correction factor to the X-ray band. They refer to \citet{Khab:2014_erosita} from where an X-ray peak luminosity of 
$1\times10^{44}$\lumunits can be derived based on the Eddington luminosity and K-correction for a $10^{6}$\msolar black hole.
\end{table*}
\end{landscape}


\bibliographystyle{aps-nameyear}   
\bibliographystyle{bstfile-doi}

\end{document}